\PassOptionsToPackage{pdftex,pdfencoding=auto,colorlinks=true, urlcolor=blue, 
                      anchorcolor=blue, citecolor=blue, filecolor=blue, 
                      linkcolor=blue, menucolor=blue, pagecolor=blue, 
                      linktocpage=true, pdfproducer=medialab, pdfa=true}{hyperref}
\pdfoutput=1
\documentclass[a4paper,11pt]{article}

\usepackage[left=2.5cm,top=4cm,right=2.5cm,bottom=3cm]{geometry}
\usepackage{amsmath,amssymb}
\usepackage{placeins}
\usepackage{float}
\usepackage{physics}
\usepackage{subfigure,epsfig,epsf,color}
\usepackage[utf8x]{inputenc}
\usepackage{amssymb,amsmath}
\usepackage{slashed}
\usepackage{graphicx}
\usepackage{graphics}
\usepackage{epstopdf}
\usepackage{subfigure}
\usepackage{array}
\usepackage{multirow}
\usepackage{amsfonts}
\usepackage{pdfcomment}
\usepackage{hyperref}
\usepackage{cite}
\usepackage{pdflscape}
\usepackage{orcidlink}
\usepackage{lipsum}
\usepackage{jcappub}
\usepackage{comment}
\usepackage{array}
\usepackage{tabularray}
\usepackage{mathrsfs}
\usepackage{comment}
\usepackage{color}
\usepackage{comment}
\usepackage{array}
\usepackage{tabularray}
\usepackage{makecell}
\usepackage{cellspace}
\usepackage{pdflscape}
\usepackage{chngpage}
\usepackage{subfigure}

\newcolumntype{M}[1]{>{\centering\arraybackslash}m{#1}}
\usepackage[T1]{fontenc}
\title{Dynamical systems analysis of Bianchi type-IX cosmology with a scalar field mediated dark sector interaction}
\author[a]{Ashmita\orcidlink{0000-0003-3252-8021},}
\author[a]{Jaturaporn Watanakampolkij\orcidlink{0009-0008-4926-0387},}
\author[a,1]{Daris Samart\orcidlink{0000-0003-4690-7591}\note{Corresponding author}}

\affiliation[a]{Khon Kaen Particle Physics and Cosmology Theory Group (KKPaCT), \\
Department of Physics, Faculty of Science, Khon Kaen University}
 
\emailAdd{ashmitarai.physics@gmail.com}
\emailAdd{jaturaporn.wa@kkumail.com}
\emailAdd{darisa@kku.ac.th}

\abstract{ 
The recent DESI DR2 results preference for a time-varying dark-energy equation of state, together with the long-standing difficulty of explaining why an initially anisotropic universe should evolve toward isotropy, motivates the study of interacting dark-sector models in anisotropic backgrounds. We investigate the dynamics of Bianchi type-IX cosmologies containing pressureless dark matter coupled to dark energy in the form of a quintessence scalar field $\phi$ with exponential potential $V(\phi)=V_0\,e^{-\lambda\phi}$. Energy exchange between the two dark components is mediated by the scalar field through the interaction term $Q=\kappa\beta\dot\phi\,\rho_{\rm DM}$, where $\kappa^2=8\pi G$ and $\beta$ is a dimensionless coupling. Because Bianchi~IX models generically recollapse, the Hubble scalar
$H$ vanishes at the moment of maximum expansion, where the expansion-normalized variables of the standard Wainwright--Ellis formalism diverge. We therefore adopt the $D$-normalized phase-space formalism of Hewitt, Uggla, and Wainwright, which compactifies the state space and remains regular through the recollapse, including the moment of maximum expansion at which $H = 0$. Together with the modified conservation equations for the two dark components, the Einstein equation reduces to a closed autonomous polynomial dynamical system, whose fixed points we classify by linear stability analysis, supplemented by center-manifold theory for the non-hyperbolic points. The fixed points shows the kinetic-dominated, coupling modified matter dominated, curvature scalar, scalar field dominated and coupled scaling families; the last is the anisotropic analogue of the Amendola scaling attractor. Restricting to the initially expanding, isotropizing trajectories, we map the region of the $(\lambda,\beta)$ plane in which anisotropic Bianchi~IX models approach an accelerating Friedmann--Robertson--Walker state. \\

}

\begin{document}
\maketitle
\flushbottom

\section{Introduction}
Over the past three decades, cosmological observations supported by the measurements of Type Ia Supernovae \cite{Riess1998, Perlmutter1999}, Cosmic Microwave Background (CMB) \cite{Planck2018}, and Baryon Acoustic Oscillations (BAO) \cite{Eisenstein2005, Alam2017} have firmly established that the universe is undergoing accelerated expansion. Within the standard model of Cosmology given by the $\Lambda\text{CDM}$ model, this late time cosmic acceleration is attributed to a positive cosmological constant $\Lambda$, while cold dark matter drives the growth of cosmic structure. Together, Dark Energy (DE) and Dark Matter (DM) constitute approximately $95\%$ of the total cosmic energy budget. However, despite successfully accounting for observational data, the standard model faces significant theoretical difficulties, notably the fine tuning of $\Lambda$ and the cosmic coincidence problem \cite{Weinberg1989, ZlatevWangSteinhardt1999, Carroll2001} concerning why the dark matter energy density ($\rho_{\text{DM}}$) and dark energy density ($\rho_{\text{DE}}$) are of the same order of magnitude at the present epoch. To address these theoretical impasses, dynamical DE models featuring a time varying equation of state (EoS) parameter $w_{\text{DE}} = p_{\text{DE}}/\rho_{\text{DE}}$ \cite{Ratra1988, Wetterich1988, CaldwellDaveSteinhardt1998, CopelandSamiTsujikawa2006} have been widely investigated. 

In this context, slowly rolling scalar fields, most prominently canonical quintessence, phantom and k essence fields, provide the standard mechanism to reproduce late time cosmic acceleration dynamically \cite{Ratra1988, Wetterich1988, Caldwell2002, CopelandSamiTsujikawa2006}. To accommodate richer dynamical phenomenology, such as crossing the phantom divide line ($w_{\text{DE}} = -1$) or driving assisted cosmic acceleration, multi scalar field frameworks including quintom and assisted quintessence models have also been extensively formulated \cite{ColeyVdH2000, Feng2005, GuoCaiZhang2005, Cai2010}. Extending this framework to include a non gravitational coupling between DM and DE provides a straightforward and physically grounded scenario \cite{Wetterich1995, Amendola1999, WangEtAl2016, BahamondeEtAl2018}. Since no fundamental gauge symmetry prevents interactions within the dark sector, energy and momentum can be continuously exchanged through an interaction term $Q$, modifying the fluid conservation equations to include a source term:
\begin{equation}
\dot\rho_{\text{DM}} + 3H\rho_{\text{DM}} = Q, \qquad \dot\rho_{\phi} + 3H(\rho_{\phi} + p_{\phi}) = -Q.
\end{equation}

The foundation for this mechanism traces back to early scalar field models featuring a time dependent cosmological term \cite{Wetterich1988}, and the interaction itself was placed on a systematic footing by Amendola \cite{Amendola2000} in the coupled quintessence scenario, where the exchange arises naturally from a scalar field non minimally coupled to the matter sector. Its appeal is twofold. First, a suitable coupling admits scaling (or ``stationary'') solutions in which the DE density tracks that of DM, so that the two remain comparable for an extended period; this alleviates the cosmic coincidence problem, which an uncoupled cosmological constant does not address \cite{Amendola2000, Zimdahl2001, GuoCaiZhang2005, ChenGongSaridakis2009}. Second, the energy drained from DM can itself sustain the accelerated expansion, so that acceleration is driven by the interaction rather than being put in by hand. Interacting DE models have since been studied extensively against cosmological data \cite{WangAbdallaAtrioBarandela2016}, and their late time behaviour is naturally analysed with dynamical systems methods. This approach reduces the cosmological field equations to an autonomous system whose fixed points and their stability classify the possible asymptotic states independently of fine tuned initial conditions \cite{CopelandSahniTsujikawa2006, BahamondeBohmerCarloni2018}. Furthermore, comprehensive reviews of dark sector interactions \cite{WangEtAl2016, BahamondeEtAl2018} alongside recent observational constraints from the Dark Energy Spectroscopic Instrument (DESI) \cite{DESI_DR2_2025, DESI_BAO_2025} have renewed interest in coupled dark sector models as compelling candidates for dynamical dark energy and effective phantom divide crossings \cite{InteractingDE_DESI_2025, YukawaCoupled_2025, DESIcoupled2025}. 

While the isotropic and spatially flat Friedmann Lema{\^i}tre Robertson Walker (FLRW) metric serves as the standard background model, large scale anomalies in the CMB such as the quadrupole octupole alignment and directional power asymmetries suggest residual departures from exact spatial isotropy \cite{deOliveiraCosta2004, SchwarzCopiHuterer2016, Planck2018Anomalies}. Spatially homogeneous but anisotropic spacetimes, classified under the Bianchi types, provide the standard general relativistic framework for probing directional expansion rates, shear evolution and cosmic isotropization mechanisms \cite{CollinsHawking1973, Wald1983, WainwrightHsu1989, IbanezVdHColey1995, vdHOlasagasti1999}. Evaluating an interacting dark sector within an anisotropic Bianchi metric therefore provides a robust method to determine whether energy exchange between dark components can damp anisotropic shear and ensure late time isotropization \cite{Bhanja2023, RathoreSingh2023, MillanoPaliathanasis2025}. 

Indeed, accounting for observed spatial uniformity within general relativity presents a longstanding theoretical challenge \cite{CollinsStewart1971, CollinsHawking1973, EllisMacCallum1969}. In foundational work, Collins and Hawking \cite{CollinsHawking1973} examined spatially homogeneous cosmologies filled with ordinary matter satisfying the strong and dominant energy conditions \cite{HawkingEllis1973, BarrowTipler1979}. Specifically, across the space of possible initial conditions defined by the expansion rate, anisotropic shear, spatial curvature and matter density, the configurations that evolve toward an isotropic state form a set of measure zero. Consequently, under decelerating matter, a randomly selected universe almost never becomes isotropic. Spatial uniformity is thus highly non generic, requiring an explicit dynamical mechanism to achieve it.

Accelerated expansion provides a robust dynamical mechanism that evades this restriction, rendering spatial uniformity a generic late time outcome. In the early universe, an inflationary phase dilutes anisotropic shear far more rapidly than matter densities, driving the geometry toward an isotropic de Sitter state as formalized by cosmic no hair theorems \cite{Wald1983, JensenSteinSchabes1986, RothmanEllis1986}. A parallel mechanism operates during the DE dominated epoch: Wald \cite{Wald1983} demonstrated that all initially expanding Bianchi models with a positive cosmological constant and matter obeying the strong and dominant energy conditions asymptotically approach de Sitter spacetime, with the singular exception of Bianchi type IX. This exception is physically profound. Bianchi IX models possess positive spatial curvature ($S^3$ topology) and can behave as closed universes. For ordinary matter with non negative pressure, Lin and Wald \cite{LinWald1989, LinWald1990} (see also \cite{BarrowGallowayTipler1986}) proved that such spacetimes inevitably reach a maximum volume and recollapse to a singularity rather than isotropizing. Because dark energy possesses negative pressure, it violates the energy conditions underpinning the recollapse theorem and can, in principle, halt contraction and drive late time expansion \cite{GoliathEllis1999, Heinzle2005, CalogeroHeinzle2009, KohliHaslam2014}. Nevertheless, whether dark energy can dominate before positive spatial curvature forces turnaround, and whether the resulting trajectory is genuinely isotropic once anisotropic shear and dark sector couplings are active, cannot be settled by no hair theorems alone.

To establish this geometric context rigorously, it is essential to recognize that Bianchi IX has historically served as the preeminent theoretical laboratory for probing the limits of general relativity. Foundational investigations trace back to Misner \cite{Misner1969}, who introduced the Mixmaster universe to explore early time homogenization, and Belinskii, Khalatnikov, and Lifshitz \cite{BKL1970, BKL1982}, who formulated the oscillatory asymptotic approach to the initial singularity. Built upon orthonormal frame classifications \cite{EllisMacCallum1969, RyanShepley1975} and generic singularity analyses \cite{BarrowTipler1979}, the chaotic nature of the Mixmaster evolution was extensively debated \cite{BurdBuricEllis1990, RughJones1990} and ultimately confirmed using invariant fractal methods by Cornish and Levin \cite{CornishLevin1997}. Global dynamics and exact attractor structures for vacuum and perfect fluid geometries were subsequently proven rigorously by Rendall \cite{Rendall1997}, Ringstr{\"o}m \cite{Ringstrom2000, Ringstrom2001}, and Heinzle and Uggla \cite{HeinzleUggla2009}, generalizing past attractor frameworks \cite{UgglaEtAl2003}. Beyond simple fluids, researchers have continuously introduced complex sources into this closed geometry to test their effects on cosmic evolution. These include magnetic fields \cite{LeBlancKerrWainwright1995, LeBlanc1997}, Vlasov and elastic matter \cite{CalogeroHeinzle2009, CalogeroHeinzle2010}, and viscous fluids \cite{KohliHaslam2014}, alongside stability analyses of the Einstein static universe against Bianchi IX perturbations \cite{BarrowEllisMaartensTsagas2003, BarrowYamamoto2012}.

To systematically determine whether a coupled dark sector can similarly halt recollapse and drive late time isotropization, the dynamical systems approach provides the standard framework to map global phase space trajectories across all evolutionary epochs. Pioneered by Wainwright, Ellis, and collaborators \cite{WainwrightHsu1989, WainwrightEllis1997}, this methodology recasts the non linear field equations into an autonomous system. In anisotropic cosmologies, standard formulations utilize expansion normalized variables scaled by the Hubble parameter $H$. While this formulation has been applied successfully to ever expanding Bianchi models, it encounters a coordinate singularity in Bianchi type IX: due to positive spatial curvature, generic type IX spacetimes reach a maximum volume where $H=0$ at turnaround, causing $H$ normalized variables to diverge. To resolve this singularity, Hewitt, Uggla, and Wainwright (HUW) introduced a regularized normalization based on the strictly positive quantity \cite{HUW1997, WainwrightEllis1997}
\begin{equation}
D = \sqrt{H^2 + \tfrac{1}{4}(n_1 n_2 n_3)^{2/3}},
\end{equation}
where $n_i$ denote the spatial structure constants of the Bianchi IX Lie algebra. Because $D > 0$ globally, normalizing by $D$ rather than $H$ maps the dynamics onto a bounded, compact phase space that remains regular and differentiable throughout expansion, turnaround ($H=0$), and subsequent recollapse ($H < 0$).

Using this HUW formalism, van den Hoogen and Olasagasti \cite{vdHOlasagasti1999} investigated vacuum and minimally coupled scalar field Bianchi IX cosmologies with an exponential potential $V(\phi) = V_0 e^{-\lambda\phi}$. They established a bifurcation threshold: for steep potentials ($\lambda^2 > 2$), every initially expanding model inevitably succumbs to positive spatial curvature and recollapses toward an anisotropic singularity, whereas for flat potentials ($\lambda^2 < 2$), an open set of initial trajectories avoids recollapse and isotropizes asymptotically toward a power law inflationary FLRW attractor. Subsequent studies extended this regularized framework to Bianchi type $\text{VII}_h$ \cite{vdHColeyIbanez1997}, closed spacetimes containing uncoupled fluids alongside scalar fields \cite{ColeyGoliath2000}, and higher dimensional brane world scenarios \cite{vdHColeyHe2003, Coley2005brane}. 

More recently, dynamical systems investigations of an interacting dark sector have been conducted for spatially flat and locally rotationally symmetric (LRS) Bianchi type I backgrounds \cite{Bhanja2023, RathoreSingh2023, MillanoPaliathanasis2025}, confirming that dark sector energy exchange can assist in isotropizing directional expansion rates. However, in Bianchi I the spatial curvature vanishes identically, removing the primary obstacle to cosmic expansion. While previous studies have analyzed either uncoupled scalar fields in Bianchi IX or coupled dark sectors in spatially flat Bianchi I geometries, a complete autonomous system treatment of a coupled dark sector within the positively curved, recollapsing Bianchi IX geometry remains absent. 

While interacting dark sector models have been extensively studied in spatially flat anisotropic backgrounds, their dynamical behavior within the positively curved, classically recollapsing Bianchi IX geometry remains unexplored. The present work bridges this gap. In this paper, we carry out a dynamical systems analysis of Bianchi type IX cosmologies containing a quintessence scalar field with an exponential potential $V(\phi) = V_0 e^{-\lambda\phi}$ coupled to a perfect fluid with equation of state $p = \omega \rho$ through the non gravitational interaction $Q = \kappa\beta\dot\phi\rho_m$. By utilizing the regularized HUW $D$ normalization, we establish how the energy exchange alters the global cosmological phase space. The resulting autonomous system, subject to the two HUW constraints, is analyzed via the linear stability of its fixed points. We identify all equilibrium states, classify their stability, and specifically map the ($\lambda,\beta$) parameter space to demonstrate that for sufficiently flat potentials, the scalar field mediated interaction not only averts cosmic recollapse but efficiently damps primordial shear, ensuring that a fully anisotropic closed universe generically isotropizes toward a stable, accelerating FRW attractor.

The paper is organized as follows. Section~\ref{sec:setup} introduces the geometric framework of Bianchi IX and derives the complete set of $D$ normalized evolution equations alongside their constraint equations. Section~\ref{sec:isotropic} isolates the isotropic sector, presenting a fixed point analysis for each value of the fluid equation of state parameter $\gamma$, and establishes the precise conditions required for late time isotropization and scaling attractors. Section~\ref{sec:anisotropic} extends this analysis to the fully anisotropic invariant sets, evaluating the stability of Kasner like and anisotropic curvature dominated points. Section~\ref{sec:comparison} places our results in theoretical context, providing a comprehensive comparative summary table that maps our findings against the existing body of work on Bianchi IX cosmologies. Finally, Section~\ref{sec:conclusions} summarizes our core conclusions and discusses their physical implications.

\section{Dynamics of the Interacting DM-DE Model}
\label{sec:setup}
Anisotropic but spatially homogeneous cosmologies provide the natural
setting in which to ask whether and how, an initially anisotropic universe can evolve toward the isotropy we observe today. They are classified by the three-dimensional Lie algebra of their spatial isometry group, a scheme that goes back to Bianchi~\cite{Bianchi1897} and was cast in its modern orthonormal-frame form by Ellis and MacCallum~\cite{EllisMacCallum1969, MacCallum1979}. Among the nine Bianchi types, type~IX is singled out by its closed $S^3$ spatial sections and positive spatial curvature: it contains the closed FLRW model as its isotropic limit, and its vacuum and perfect-fluid dynamics display the rich oscillatory ``Mixmaster'' behaviour near the singularity studied by Misner~\cite{Misner1969} and by Belinskii, Khalatnikov and Lifshitz~\cite{BKL1970}. The same positive curvature that generates this complexity also allows type~IX models to reach a moment of maximum expansion and recollapse, which is what makes their late-time isotropization a non-trivial question and motivates the phase-space treatment developed below. In this section we set up the geometry of the Bianchi~IX model, introduce the interacting dark-matter and dark-energy content, and recast the Einstein equations as a regular autonomous dynamical system.

\subsection{Cosmological model setup}\label{sec:geometry}
In the Bianchi classification, type~IX is the only one whose spatial sections carry the topology of a 3-sphere ($S^3$), which makes it the direct anisotropic generalization of the closed FLRW model~\cite{EllisMacCallum1969, WainwrightEllis1997}. In this work the geometry is represented by the line element, written in terms of the basis 1-forms $\omega^i$ (the Maurer--Cartan forms of $SU(2)$),
\begin{equation}
ds^2 = -dt^2 + a^2(t)(\omega^1)^2 + b^2(t)(\omega^2)^2 + c^2(t)(\omega^3)^2,
\label{eq:lineelement}
\end{equation}
where $a(t),b(t),c(t)$ are the directional scale factors, so that the expansion proceeds at a different rate along each principal axis and the common limit $a=b=c$ returns the closed FLRW metric. In the Euler-angle coordinates $(\psi,\theta,\phi)$ on $S^3$, the basis 1-forms are
\begin{align}
\omega^1 &= d\psi + \cos\theta \, d\phi, \label{eq:w1}\\
\omega^2 &= \cos\psi \, d\theta + \sin\psi \sin\theta \, d\phi, \label{eq:w2}\\
\omega^3 &= \sin\psi \, d\theta - \cos\psi \sin\theta \, d\phi. \label{eq:w3}
\end{align}
They satisfy the Maurer-Cartan relations, written most simply in terms of the dual frame vectors $e_i$ as
\begin{equation}
[e_j,e_k]=\varepsilon_{ijk}\,e_i,
\label{eq:MC}
\end{equation}
whose totally antisymmetric structure constants $\varepsilon_{ijk}$ are those of the $SU(2)$ algebra and identify the group as Bianchi type~IX.

\noindent The mean expansion rate is the Hubble scalar
\begin{equation}
H=\frac13\left(\frac{\dot a}{a}+\frac{\dot b}{b}+\frac{\dot c}{c}\right),
\label{eq:hubble}
\end{equation}
and the anisotropy of the expansion is parameterized through the Misner shear variables $(\sigma_+,\sigma_-)$ defined by
\begin{equation}
\frac{\dot a}{a}-H=-2\sigma_+,\qquad
\frac{\dot b}{b}-H=\sigma_++\sqrt3\,\sigma_-,\qquad
\frac{\dot c}{c}-H=\sigma_+-\sqrt3\,\sigma_-.
\label{eq:shear}
\end{equation}
Here $\sigma_+$ measures the anisotropy between the $\omega^1$ axis and the average of the other two, while $\sigma_-$ measures the anisotropy between the $\omega^2$ and $\omega^3$ axes; the magnitude of the anisotropy is the shear scalar $\sigma^2=\sigma_+^2+\sigma_-^2$, which vanishes precisely in the isotropic limit $a=b=c$.

\begin{table}[t]
\centering
\renewcommand{\arraystretch}{1.55}
\addtolength{\tabcolsep}{-1.8pt}
\begin{tabular}{llc l}
\hline

\hline
Class & Type & \makecell{Structure constants \\ $(n_1, n_2, n_3)$} & \makecell{Spatial Geometry} \\
\hline

\hline
\multirow{6}{*}{A ($a_i=0$)}
 & I       & $(0, 0, 0)$ & \makecell{Pure Kasner/shear (Flat FLRW limit)} \\
 \hline
 & II      & $(+, 0, 0)$ & \makecell{Anisotropic flat \\
 (single non-zero spatial curvature term) }\\  \hline

 & VI$_0$  & $(0, +, -)$ & \makecell{Anisotropic open \\ (two opposite-sign curvature terms)}\\  \hline

 & VII$_0$ & $(0, +, +)$ & \makecell{Anisotropic flat \\ (two same-sign curvature terms; Flat FLRW limit)} \\  \hline

 & VIII    & $(-, +, +)$ & \makecell{Anisotropic open \\ (mixed-sign spatial curvature terms)} \\  \hline

 & IX      & $(+, +, +)$ & \makecell{Spatially closed $\leftarrow$ \emph{the paper's case}} \\
\hline

\hline
\end{tabular}
\caption{This table presents the Bianchi classification of spatially homogeneous spacetimes in the Ellis--MacCallum framework \cite{EllisMacCallum1969}. Class~A ($a_i=0$) unimodular models admit the $D$-normalized autonomous formulation used in this work. Type~IX is unique in possessing three positive structure constant eigenvalues ($n_1, n_2, n_3 > 0$), corresponding to spatially closed geometries ($S^3$ topology) with the closed FLRW ($k=+1$) model as its isotropic limit, whereas Types~I, VII$_0$, V, and VII$_h$ contain the flat ($k=0$) and open ($k=-1$) FLRW limits. Non unimodular Class~B models  where $a_i \neq 0$ are not mentioned in this table because it lies outside the scope of this work, as Bianchi~IX belongs strictly to Class~A.}
\label{tab:bianchi}
\end{table}

In the Lie algebra classification of spatially homogeneous spacetimes, non zero structure constants provide the spatial hypersurfaces with intrinsic curvature even in the presence of isotropic expansion. Within the Ellis MacCallum decomposition \cite{EllisMacCallum1969}, the structure constants are split into a symmetric matrix $n_{ab}$ accompanied by a trace vector $a_i$. Spatially homogeneous cosmologies divide naturally into unimodular Class A models ($a_i = 0$) and non unimodular Class B models ($a_i \neq 0$). Because Bianchi IX belongs strictly to Class A, the diagonal $n_i$ based formulation and the $D$ normalization employed in this work apply exclusively to unimodular geometries. In an orthonormal frame, the spatial curvature tensor for Class A models reduces to $\text{diag}(n_1, n_2, n_3)$, where $n_i(t)$ denote the spatial frame commutation functions that scale with directional scale factors as $n_1 \propto a / (b c)$ (and cyclic permutations). Within Class A, the spatial geometry is uniquely fixed by the number of vanishing eigenvalues and the sign pattern of the non zero $n_i$: Type I ($n_i = 0$) is spatially flat, Types $\text{VI}_0$ and $\text{VII}_0$ feature one vanishing eigenvalue, Type VIII exhibits a mixed sign signature, and Type IX is unique in possessing three strictly positive eigenvalues ($n_i > 0$). This all positive signature establishes the compact $S^3$ spatial topology of Bianchi IX, singling it out as the general relativistic anisotropic generalization of the closed FLRW model. Crucially, this structure carries a direct dynamical consequence: the positive definite nature of $n_i$ ensures that the term $(n_1 n_2 n_3)^{2/3}$ appearing in the $D$ normalization \eqref{eq:Ddef} remains strictly positive throughout the evolution, which is precisely what keeps the phase space regular through the moment of maximum expansion, as discussed in Section~\ref{sec:Dnorm}.

\noindent The spatial Ricci tensor components are
\begin{align}
    {}^{(3)}R^1_1&=\frac{1}{2}(n_1^2-n_2^2-n_3^2+2n_2n_3) \nonumber \\
    {}^{(3)}R^2_2&=\frac{1}{2}(n_2^2-n_1^2-n_3^2+2n_1n_3) \\
    {}^{(3)}R^3_3&=\frac{1}{2}(n_3^2-n_1^2-n_2^2+2n_1n_2) \nonumber
\end{align}
In differential geometry, any symmetric tensor can be decomposed into two parts: its trace and its trace-free part. For the spatial Ricci tensor, this decomposition is
\begin{equation*}
    {^{(3)}}R^i_j=S^i_j+\frac{1}{3}{}^{(3)}R\,\gamma^i_j
\end{equation*}
Rearranging this definition, the diagonal components simplify directly to
\begin{equation*}
    S_i^i={}^{(3)}R^i_i-\frac{1}{3}{}^{(3)}R
\end{equation*}

\noindent
The evolution of the universe is described by the Raychaudhuri equation for the Hubble parameter $(H)$, the evolution of the trace-free shear tensor components ($\sigma_+$ and $\sigma_-$), and the evolution of the spatial structure constraints $(n_1, n_2, n_3)$,
\begin{align}
    \dot{H}&=-H^2-\frac{2}{3}(\sigma_{+}^2+\sigma_{-}^2)-\frac{1}{6}(\rho_{\text{tot}}+3p_{\text{tot}}),\\
    \dot{\sigma}_{\pm}&=-3H\sigma_{\pm}-{}^{(3)}S_{\pm},\\
    \frac{\dot{n}_1}{n_1}&=-H-4\sigma_+, \\
    \frac{\dot{n}_2}{n_2}&=-H+2\sigma_++2\sqrt{3}\sigma_-, \\
    \frac{\dot{n}_3}{n_3}&=-H+2\sigma_+-2\sqrt{3}\sigma_-.
\end{align}
The system is bound by the generalized Friedmann equation, which acts as the Hamiltonian constraint connecting expansion, matter, shear, and geometry
\begin{equation}
    3H^2=\rho_{\text{tot}}+\sigma_{+}^2+\sigma_{-}^2-\frac{1}{2}{}^{(3)}R.
\end{equation}
The spatial 3-Ricci scalar $({}^{(3)}R)$ and the trace-free spatial Ricci tensor components $({}^{(3)}S_{\pm})$ are entirely determined by the geometric structure constants of the Bianchi IX group,
\begin{align}
    {}^{(3)}R&=-\tfrac12\!\left[\,n_1^2+n_2^2+n_3^2 -2\left(n_1n_2+n_2n_3+n_3n_1\right)\right],
    \label{eq:3curv}\\
    {}^{(3)}S_{+}&=\frac{1}{6}\left[(n_2-n_3)^2-n_1(2n_1-n_2-n_3)\right],\\
    {}^{(3)}S_{-}&=\frac{\sqrt{3}}{6}(n_2-n_3)(-n_1+n_2+n_3).
\end{align}
which in the isotropic limit $n_1=n_2=n_3$ reduces to the positive spatial curvature ($k>0$) of the closed FLRW model. It is this positive curvature that eventually halts the expansion and drives the recollapse discussed above and its anisotropic part sources the shear, as we make explicit in the evolution equations below.

\subsection{Matter content and the dark-sector interaction}
\label{subsec:matter}

The matter content consists of a quintessence scalar field $\phi$ with exponential potential
\begin{equation}
V(\phi) = V_0\,e^{-\lambda\phi},
\end{equation}
representing the dark energy, coupled to a barotropic perfect fluid of dark matter with energy density $\rho$ and pressure $p = (\gamma - 1)\rho$, where the equation-of-state parameter $\gamma \in [1,2]$ is left free at the level of the dynamical system and will be specified later to dust ($\gamma=1$), radiation ($\gamma=4/3$) and stiff fluid ($\gamma=2$) while studying the fixed points of the dynamical system. The total energy momentum tensor is the sum $T^{\text{tot}}_{ab} = T^\phi_{ab} + T^{\text{DM}}_{ab}$ of the scalar-field and fluid contributions. Following Amendola~\cite{Amendola1999}, we take the energy exchange between the two components to be mediated by the scalar field through the interaction
\begin{equation}
\label{eq:interaction}
Q = \kappa\beta\,\dot\phi\,\rho_{\text{DM}},
\end{equation}
where $\beta$ is a dimensionless coupling constant. The modified conservation equations read
\begin{align}
\label{eq:fluid-cont}
\dot\rho_{\text{DM}} + 3H\gamma\,\rho_{\text{DM}} &= +Q, \\
\label{eq:KG}
\ddot\phi + 3H\dot\phi + \frac{dV}{d\phi} &= -\frac{Q}{\dot\phi},
\end{align}
so that the total energy density is covariantly conserved while the individual components exchange energy at the rate $Q$. The sign convention is such that $\beta < 0$ corresponds to energy transfer from dark matter to dark energy. Substituting the interaction \eqref{eq:interaction} into the Klein Gordon equation \eqref{eq:KG} removes the explicit $1/\dot\phi$ factor, leaving a source term $-\kappa\beta\rho_{\rm DM}$ that acts as an effective, density-dependent force on the scalar field. The coupling therefore modifies the scalar-field dynamics only while a non-negligible dark-matter density is present, and it switches off once the fluid has redshifted away.

\subsection{$D$-normalization and dimensionless variables}
\label{sec:Dnorm}

Because Bianchi~IX models with ordinary matter obeying the strong and dominant energy conditions generically recollapse~\cite{LinWald1990}, the Hubble scalar $H$ vanishes at the moment of maximum expansion and the standard expansion normalization breaks down. We therefore use the $D$-normalization of Hewitt, Uggla, and Wainwright~\cite{HUW1997}, in which dimensionless variables are obtained by dividing by the strictly positive quantity
\begin{equation}
\label{eq:Ddef}
D \equiv \sqrt{H^2 + \tfrac{1}{4}(n_1 n_2 n_3)^{2/3}}.
\end{equation}
The combination of $H^2$ with the structure-constant term ensures $D > 0$ throughout the evolution of any Bianchi~IX model: at the moment of maximum expansion $H = 0$ but $n_1 n_2 n_3 > 0$, while at the singularity boundary $n_i \to 0$ but $H^2 > 0$.

We introduce a new dimensionless time variable $\tau$ defined by $d\tau/dt = D$, and denote $\tau$-derivatives by a prime: 
$X' \equiv dX/d\tau = \dot X / D$. The complete set of $D$-normalized variables for the geometric, scalar-field and fluid sectors is
\begin{align}
\label{eq:dimvar-geom}
\tilde H &= \frac{H}{D}, &
\tilde\Sigma_+ &= \frac{\sigma_+}{D}, &
\tilde\Sigma_- &= \frac{\sigma_-}{D}, &
\tilde N_i &= \frac{n_i}{D}, \\
\label{eq:dimvar-scalar}
\tilde X &= \frac{\kappa\,\dot\phi}{\sqrt{6}\,D}, &
\tilde Y &= \frac{\kappa\,\sqrt{V}}{\sqrt{3}\,D}, &
\tilde\Omega_{\text{DM}} &= \frac{\kappa^2 \rho_{\text{DM}}}{3 D^2},
\end{align}
where $\kappa^2 = 8\pi G$. The variables $\tilde X^2$ and 
$\tilde Y^2$ represent the kinetic and potential contributions of 
the scalar field to the Friedmann constraint and satisfy 
$\tilde X^2 + \tilde Y^2 = \tilde\Omega_\phi$, with 
$\tilde\Omega_\phi = \kappa^2\rho_\phi/(3D^2)$.
The sign of $\tilde H$ distinguishes expanding ($\tilde H>0$) from
contracting ($\tilde H<0$) phases and the turnaround at maximum expansion where $\tilde H=0$, the $D$-normalization reaches at finite $\tau$ without any variable diverging. In this way a complete Bianchi~IX evolution: expansion, turnaround and recollapse, is mapped onto a bounded trajectory in the compact state space, and the standard machinery of fixed-point analysis becomes applicable.

\subsection{Evolution equations and constraints}
\label{subsec:evolution}

Two algebraic constraints must hold throughout the evolution. The first follows directly from the definition of $D$ in 
Eq.~\eqref{eq:Ddef},
\begin{equation}
\tilde H^2+\tfrac14(\tilde N_1\tilde N_2\tilde N_3)^{2/3}=1,
\label{eq:constraintD}
\end{equation}
and shows that the state space is a compact algebraic surface with $-1\le\tilde H\le1$. The second is the $D$-normalized Friedmann constraint, obtained from the $tt$ component of the Einstein field equations,
\begin{equation}
\tilde H^2=\tilde\Sigma_+^2+\tilde\Sigma_-^2+\tilde X^2+\tilde Y^2
+\tilde\Omega_{\rm DM}
+\tfrac1{12}\!\left[\left(\tilde N_1^2+\tilde N_2^2+\tilde N_3^2\right)
-2\left(\tilde N_1\tilde N_2+\tilde N_2\tilde N_3+\tilde N_1\tilde N_3\right)\right],
\label{eq:friedmann}
\end{equation}
which gives $\tilde H^2$ as the sum of shear, scalar-field, fluid and spatial-curvature contributions.

The two constraints \eqref{eq:constraintD} and \eqref{eq:friedmann} are preserved by the flow and reduce the number of independent variables: \eqref{eq:constraintD} fixes one of the $\tilde N_i$ in terms of the other two and $\tilde H$, while the Friedmann constraint \eqref{eq:friedmann} determines $\tilde\Omega_{\rm DM}$ algebraically in terms of the remaining variables. The fluid density $\tilde\Omega_{\rm DM}$ is therefore not an independent degree of freedom, and its evolution equation \eqref{eq:OmegaDM} below serves as a consistency check on the reduced system rather than as an additional equation. The autonomous system of evolution equations follows from those of van den Hoogen and Olasagasti~\cite{vdHO1999} for the gravitational and scalar-field sectors, modified by the interaction terms in Eqs.~\eqref{eq:fluid-cont}-\eqref{eq:KG}, 
together with the continuity equation~\eqref{eq:fluid-cont} for the fluid:
\begin{align}
\label{eq:eom-H}
\tilde H' &= \tilde q\,(\tilde H^2 - 1), \\
\label{eq:eom-Sigma-plus}
\tilde\Sigma_+' &= (\tilde q - 2)\,\tilde H\,\tilde\Sigma_+ - \tilde S_+, \\
\label{eq:eom-Sigma-minus}
\tilde\Sigma_-' &= (\tilde q - 2)\,\tilde H\,\tilde\Sigma_- - \tilde S_-, \\
\label{eq:eom-N1}
\tilde N_1' &= \tilde N_1\,(\tilde H\,\tilde q - 4\tilde\Sigma_+), \\
\label{eq:eom-N2}
\tilde N_2' &= \tilde N_2\,(\tilde H\,\tilde q + 2\tilde\Sigma_+ + 2\sqrt{3}\,\tilde\Sigma_-), \\
\label{eq:eom-N3}
\tilde N_3' &= \tilde N_3\,(\tilde H\,\tilde q + 2\tilde\Sigma_+ - 2\sqrt{3}\,\tilde\Sigma_-), \\
\label{eq:eom-X}
\tilde X' &= (\tilde q - 2)\,\tilde H\,\tilde X + \sqrt{\tfrac{3}{2}}\,\lambda\,\tilde Y^2 - \sqrt{\tfrac{3}{2}}\,\beta\,\tilde\Omega_{\text{DM}}, \\
\label{eq:eom-Y}
\tilde Y' &= \tilde Y\bigl[(1 + \tilde q)\,\tilde H - \sqrt{\tfrac{3}{2}}\,\lambda\,\tilde X\bigr] 
\end{align}
and the curvature terms $\tilde S_\pm$, which source the shear, are
\begin{align}
\label{eq:S-plus}
\tilde S_+ &= \frac{1}{6}\left(\tilde N_2 - \tilde N_3 \right)^2 - \frac{1}{6}\,\tilde N_1\,\left(2\tilde N_1 - \tilde N_2 - \tilde N_3\right), \\
\label{eq:S-minus}
\tilde S_- &= \frac{\sqrt{3}}{6}\,\left(\tilde N_2 - \tilde N_3\right) \left(-\tilde N_1 + \tilde N_2 + \tilde N_3\right).
\end{align}
Equation~\eqref{eq:eom-H} shows that the surfaces $\tilde H=\pm1$ are
invariant: a model that is expanding cannot cross to $\tilde H=1$ in finite $\tau$, so the sign of $\tilde H$ is conserved along each orbit and the expanding and contracting sectors are dynamically separated. The shear variables $\tilde\Sigma_\pm$ in \eqref{eq:eom-Sigma-plus}--\eqref{eq:eom-Sigma-minus} are driven by the anisotropic spatial curvature through the source terms $\tilde S_\pm$; when the curvature is isotropic these sources vanish and the shear decays, which is the mechanism by which a trajectory can approach an isotropic state. The scalar-field equations \eqref{eq:eom-X}--\eqref{eq:eom-Y} carry the coupling explicitly through the $\beta$ term in $\tilde X'$, which is the only place the dark-sector interaction enters the gravitational sector.

The $D$-normalized Raychaudhuri equation is
\begin{equation}
    \frac{\dot{H}}{D^2}=-\tilde{H}^2-\left[2\tilde{\Sigma}_+^2+2\tilde{\Sigma}_-^2+2\tilde{X}^2-\tilde{Y}^2+\frac{1}{2}(3\gamma-2)\tilde{\Omega}_{\text{DM}}\right],
\end{equation}
The entire bracketed expression is defined as the dimensionless deceleration parameter

\begin{equation}
\label{eq:deceleration}
\tilde q = 2\tilde\Sigma_+^2 + 2\tilde\Sigma_-^2 + 2\tilde X^2 - \tilde Y^2 + \frac{1}{2} \left(3\gamma - 2 \right)\,\tilde\Omega_{\text{DM}},
\end{equation}

\begin{equation}
    \tilde{q}=\tilde{H}^2q,
    \label{eq:deceleration-relation}
\end{equation}

and
\begin{equation}
\label{eq:OmegaDM}
\tilde\Omega_{\text{DM}}'= \tilde\Omega_{\text{DM}}\Bigl[2(1 + \tilde q)\,\tilde H - 3\gamma\,\tilde H + \sqrt{6}\,\beta\,\tilde X\Bigr]
\end{equation}
In an anisotropic universe, the shear acts geometrically like an invisible fluid composed of stiff matter, which heavily decelerates expansion. To find the equation of state that dictates the volume expansion, we use the standard cosmological relation
\begin{equation}
    \omega_{\text{tot}}=\frac{2q-1}{3}.
\end{equation}
As we shown in the above, the physical deceleration parameter $q$ is related to our dimensionless deceleration parameter by Equation~\eqref{eq:deceleration-relation},
\begin{equation}
    \omega_{\text{tot}}=\frac{(\gamma-1)\tilde{\Omega}_{\text{DM}}+\tilde{\Sigma}^2+\tilde{X}^2-\tilde{Y}^2-\frac{1}{3}\tilde{\Omega}_{K}}{\tilde{\Omega}_{\text{DM}}+\tilde{\Sigma}^2+\tilde{X}^2+\tilde{Y}^2+\tilde{\Omega}_K},
\end{equation}
where $\tilde{\Sigma}^2=\tilde{\Sigma}_+^2+\tilde{\Sigma}_-^2$ and $\tilde{\Omega}_K=\frac{1}{12}\left[\left(\tilde N_1^2+\tilde N_2^2+\tilde N_3^2\right)
-2\left(\tilde N_1\tilde N_2+\tilde N_2\tilde N_3+\tilde N_1\tilde N_3\right)\right]$.

\noindent Equations~\eqref{eq:eom-H}--\eqref{eq:eom-Y}, together with the constraints~\eqref{eq:constraintD}--\eqref{eq:friedmann}, form a closed autonomous polynomial dynamical system on a compact state space, suitable for the fixed-point analysis carried out in Section~\ref{sec:isotropic}.

\section{Fixed points of the isotropic sector}\label{sec:isotropic}

We begin by restricting the dynamical system to its isotropic invariant subspace, defined by vanishing shear ($\tilde\Sigma_+=\tilde\Sigma_-=0$) and equal spatial curvature variables ($\tilde N_1=\tilde N_2=\tilde N_3$). Once the shear of an initially anisotropic Bianchi~IX model decays, the model isotropizes and its trajectory approaches one of these fixed points at late times given in Table \ref{table:gamma}. Throughout this section we work in the physically relevant regime $\beta<0$, for which energy is transferred from DM to DE, and we keep the barotropic index $\gamma$ of the DM fluid general, so that the results cover dust ($\gamma=1$), radiation ($\gamma=4/3$) and the stiff case ($\gamma=2$). Here we consider an expanding universe, $\tilde H>0$, since we are interested in its late-time behaviour. This also fixes $\lambda>0$: at $P_4$ one has $\tilde H=\lambda/\sqrt2$, so $\tilde H>0$ requires $\lambda>0$. 
\noindent In this section, we study the isotropization of the Bianchi IX model for an interacting dark matter and dark energy setup, it admits a discrete set of fixed points that organize into five physical families, distinguished by the dominant energy component and by whether the spatial curvature is dynamically active. Each of these listed fixed point has a contracting counterpart i.e. for $\tilde H\to-\tilde H$ which represents the contracting phase of the universe, since we are interested in the late-time behaviour of an expanding universe, we list and discuss only the expanding representatives.

Before specifying the dust case, we read off the general structure of Table~\ref{table:gamma}. The kinetic energy dominated points $P_1$ and $P_2$, the curvature scalar point $P_4$ and the scalar field dominated point $P_5$ all have $\tilde\Omega_m=0$, so they are unchanged by the choice of fluid: their coordinates, $\tilde q$ and $\omega_{tot}$ are the same for dust, radiation or a stiff fluid. The barotropic index enters only through the two points that carry matter density, $P_3$ and $P_6$. At $P_3$, the whole point tracks $\gamma$ through the combination $\gamma-2$. When the coupling is switched off, $\beta\to0$, it reduces to ordinary fluid domination, $\omega_{tot}=\gamma-1$ and $\tilde q=(3\gamma-2)/2$, recovering the matter values for dust ($\omega_{tot}=0$, $\tilde q=\tfrac12$) and the radiation values for $\gamma=4/3$ ($\omega_{tot}=\tfrac13$, $\tilde q=1$); the coupling then shifts both by $\beta$-dependent terms. Its existence window in the coupling, $\beta^2\le\tfrac32(\gamma-2)^2$, is widest for dust
($\beta^2\le\tfrac32$), narrows for radiation ($\beta^2\le\tfrac23$), and closes as $\gamma\to2$, where the fluid becomes stiff and behaves like the kinetic energy of the scalar field itself, so $P_3$ no longer stands apart as a matter point.

To determine whether these configurations can act as genuine late-time attractors, we must evaluate their stability at each fixed point. In order to do that, we linearize the autonomous system around each fixed point $\mathbf{x}_0$ and compute the eigenvalues $\mu_i$ of the associated Jacobian matrix $\mathcal{J}$. Because the isotropic subspace is embedded within the broader eight-dimensional state space, a fixed point acts as a stable late-time attractor only if the real parts of all eight eigenvalues are strictly negative, ensuring that both scalar density fluctuations and geometric anisotropies (shear and curvature) decay simultaneously. Table~\ref{table:2} presents these eigenvalues evaluated for pressureless dust ($\gamma=1$).

\begin{table}[htbp]
\addtolength{\tabcolsep}{-2.6pt}
\begin{adjustwidth}{-0.05cm}{-1cm}
    \centering
    \begin{tabular}{|Sc|Sc|Sc|Sc|Sc|Sc|Sc|Sc|}
    \hline

    \hline
    Point & $\tilde{H}$ & $\tilde{N}_i$ & $\tilde{X}$ & $\tilde{Y}$ & $\tilde{\Omega}_m$ & $\tilde{q}$ & $\omega_{tot}$ \\ \hline

    \hline
    $P_{1}$ & $1$ & $0$ & $-1$ & $0$ & $0$ & $2$ & $1$ \\ \hline
    $P_{2}$ & $1$ & $0$ & $1$ & $0$ & $0$ & $2$ & $1$ \\ \hline
    $P_{3}$ & $1$ & $0$ & $-\frac{\sqrt{\frac{2}{3}}\beta}{\gamma-2}$  & $0$ & $1-\frac{2 \beta ^2}{3(\gamma-2)^2}$ & $-1-\frac{\beta^2}{\gamma-2}+\frac{3\gamma}{2}$ & $-1+\frac{2\beta^2}{6-3\gamma}+\gamma$ \\ \hline
    $P_{4}$ & $\frac{\lambda}{\sqrt{2}}$ & $\sqrt{2}{\sqrt{2-\lambda^2}}$ & $\frac{1}{\sqrt{3}}$ & $\sqrt{\frac{2}{3}}$ & $0$ & $0$ & $-\frac{1}{3}$ \\ \hline
    $P_{5}$ & $1$ & $0$ & $\frac{\lambda}{\sqrt{6}}$ & $\sqrt{1-\frac{\lambda^2}{6}}$ & $0$ & $\frac{\lambda^2-2}{2}$ & $\frac{\lambda^2-3}{3}$ \\ \hline
    $P_{6}$ & $1$ & $0$ & $\sqrt{\frac{3}{2}}\left(\frac{\gamma}{\beta+\lambda}\right)$ & $\frac{\sqrt{\beta(\beta+\lambda)-\frac{3}{2}(\gamma-2)}}{\beta+\lambda}$ & $\frac{-3+\lambda(\beta+\lambda)}{(\beta+\lambda)^2}$ & $\frac{\lambda-2\beta}{2(\beta+\lambda)}$ & $-\frac{\beta}{\beta+\lambda}$ \\ \hline

    \hline
    \end{tabular}
    \caption{Fixed points of the isotropic sector of the coupled Bianchi type-IX system for an
    arbitrary barotropic index $\gamma$ of the dark-matter fluid, together with
    the deceleration parameter $\tilde q$ and the total effective equation of
    state $\omega_{tot}$ for the expanding case. The kinetic energy dominated points $P_1$, $P_2$, the curvature-scalar point $P_4$ and the scalar field dominated point $P_5$ have vanishing matter density and are independent of $\gamma$; the
    coupling-modified matter point $P_3$ and the coupled scaling point $P_6$
    depend on $\gamma$ as shown.}
    \label{table:gamma}
\end{adjustwidth}
\end{table}

\subsection{Physical interpretation and stability of the fixed points}
\label{subsec:interpretation}

\begin{itemize}
    \item \textbf{Kinetic dominated points ($P_1$ and $P_2$):}
    At these points the scalar field carries only kinetic energy which makes the matter density vanish and the field behaves as a stiff fluid. The spatial curvature is dynamically inactive, so the geometry is effectively flat and isotropic. Both points share the $\omega_{tot}=1$ and the strongly decelerating value $\tilde q=2$. They are the anisotropic counterparts of the massless scalar field solutions and describe an early time, kinetic energy dominated phase (kination phase). Both $P_1$ and $P_2$ share four strictly positive eigenvalues $\{4,2,2,2\}$. The coupling-dependent eigenvalues further dictate their specific unstable nature: at $P_1$, $3-\sqrt{6}\beta$ and $\sqrt{3/2}\lambda+3$ are strictly positive for $\beta < 0$, making it an early-time source (past attractor). At $P_2$, $\sqrt{6}\beta+3$ becomes negative for $\beta < -\sqrt{3/2}$, converting it into a saddle. In all cases, both points possess unstable directions and describe early-time, kination phase. 

    \item \textbf{Coupling-modified matter point ($P_3$):} This configuration represents an intermediate matter-dominated era structurally modified by the dark-sector interaction. The matter density in this case is $\tilde\Omega_m=1-2\beta^2/3$, while the remaining energy budget is carried by the scalar field's kinetic energy ($\tilde X=-\sqrt{2/3}\,\beta$) dragged along by the fluid coupling. This is the Bianchi~IX analogue of the coupled quintessence matter-dominated epoch~\cite{Amendola2000}. Operating entirely as a genuine decelerating era, its stability analysis reveals a threefold positive eigenvalue $\beta^2 + \tfrac{1}{2}$ alongside a threefold negative eigenvalue $\beta^2 - \tfrac{3}{2}$ across its physical range ($\beta^2 \le \tfrac{3}{2}$). The coexistence of these signs confirms $P_3$ is a saddle point, guaranteeing a viable,transient matter dominated phase that naturally evolves toward a late-time accelerating attractor. It is physical only for $\beta^2\le3/2$, where $\tilde\Omega_m\ge0$; in the uncoupled limit $\beta\to0$ the kinetic energy switches off, $\tilde\Omega_m\to1$, and standard pressureless matter domination ($\omega_{tot}\to0$) is recovered.

    \item \textbf{Curvature scalar point ($P_4$):} 
    Here, spatial curvature is dynamically active and balances the scalar field while the matter density vanishes. Existing only for flat potentials ($\lambda^2<2$), this state represents a curvature scaling  \cite{vdHoogenColeyIbanez1997, ColeyGoliath2000} regime acting as the precise borderline between accelerated and decelerated expansion  ($\omega_{tot}=-1/3, \tilde q=0$). For $\lambda^2<2$, where $P_4$ exists, the eigenvalue $(\sqrt{8-3\lambda^2}-\lambda)/\sqrt2$ is strictly positive. The point therefore has an unstable direction and is a saddle; the eigenvalues built from $\sqrt{9\lambda^2-16}$ form complex-conjugate pairs with negative real part (a stable spiral in those directions), and there is one marginal direction with zero eigenvalue, but the single positive eigenvalue is enough to exclude $P_4$ as an attractor. Consequently, the closed Bianchi~IX geometry does not settle into this curvature-dominated configuration; it acts purely as a transient saddle that repels trajectories before late-time isotropization.

    \item \textbf{Scalar field dominated point ($P_5$):}
    At this fixed point the scalar field dominates completely, with both kinetic and potential energy present while the matter density vanishes and the curvature remains inactive. Familiar from standard exponential quintessence \cite{Copeland1998, CopelandSamiTsujikawa2006}, this point drives cosmic acceleration with $\omega_{tot} = (\lambda^2-3)/3$. It acts as a stable node and the sole late-time attractor of the isotropic sector provided $\lambda^2 < 2$. In this parameter regime, seven of its eigenvalues: $\tfrac{1}{2}(\lambda^2-6)$ and $\tfrac{1}{2}(\lambda^2-2)$ (each with multiplicity three), plus $\lambda^2-2$, are strictly negative. The remaining eigenvalue, $\lambda^2+\beta\lambda-3$, is also negative for $\beta < 0$. Because the shear and curvature modes are among the decaying directions, an initially anisotropic Bianchi IX universe in this range isotropizes into an accelerating, spatially flat state with $\omega_{\text{tot}} = (\lambda^2-3)/3 < -\tfrac{1}{3}$. For $2 < \lambda^2 < 6$, $P_5$ turns into a saddle as $\tfrac{1}{2}(\lambda^2-2)$ changes sign; hence, the boundary of stability coincides precisely with the onset of cosmic acceleration.

    \item \textbf{Coupled scaling point ($P_6$):}
    At this point both the scalar field and the matter remain dynamically important and their energy densities maintain a fixed ratio which refers to a scaling solution. The curvature is inactive here and the $\omega_{tot}=-\beta/(\beta+\lambda)$ governed jointly by the potential slope and the interaction coupling~\cite{Amendola2000, CopelandSamiTsujikawa2006}. The point is physical, with $\tilde\Omega_m\in[0,1]$, when $\lambda(\beta+\lambda)\ge3$ and $\beta(\beta+\lambda)+3/2\ge0$, and it accelerates, $\tilde q<0$, for $\lambda<2\beta$. Within this existence region the eigenvalues $1-3\beta/(\beta+\lambda)$ and the threefold $(\lambda-2\beta)/[2(\beta+\lambda)]$ are positive, while the remaining eigenvalues including the complex conjugate pair built from the square root, have negative real part. The mixed signs make $P_6$ a saddle throughout its physical existence range. Although the comparable dark matter and dark energy densities at a scaling solution would in principle be attractive from the point of view of the cosmic coincidence problem, in the present model with $\beta<0$ this scaling point is neither accelerating nor stable, and the universe does not end on it. The late-time attractor is instead the scalar-field dominated point $P_5$.

\end{itemize}

\subsection{Evolution plots for the isotropic case}

\begin{figure}[htbp]
\begin{minipage}[b]{1\linewidth}
\centering
\subfigure[\label{fig:Evolb01}]{\includegraphics[width=0.49\linewidth]{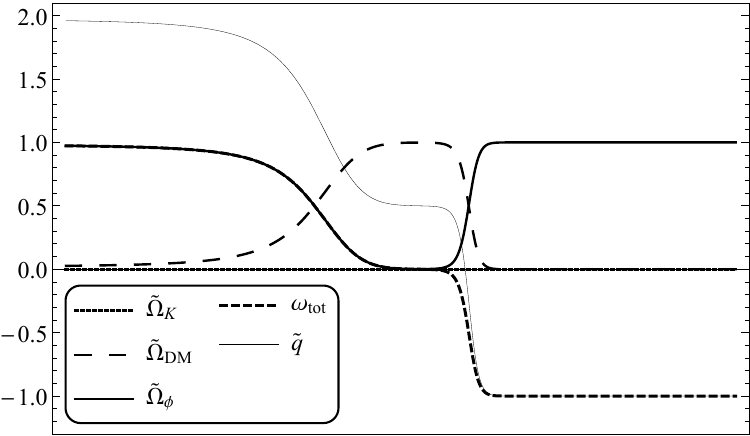}}
\subfigure[\label{fig:Evolb02}]{\includegraphics[width=0.49\linewidth]{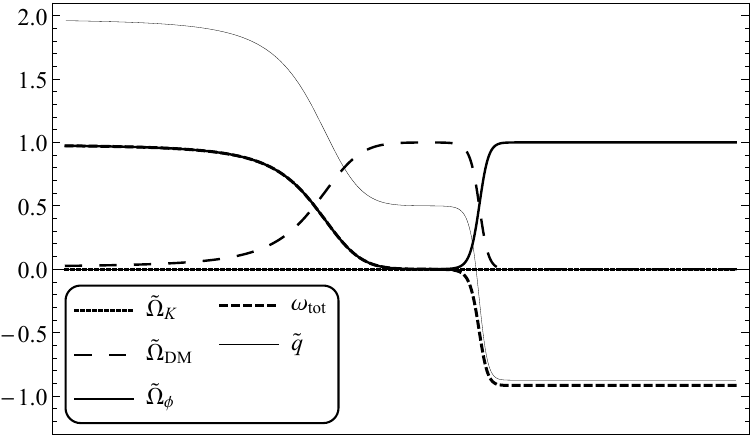}}
\subfigure[\label{fig:Evolb03}]{\includegraphics[width=0.49\linewidth]{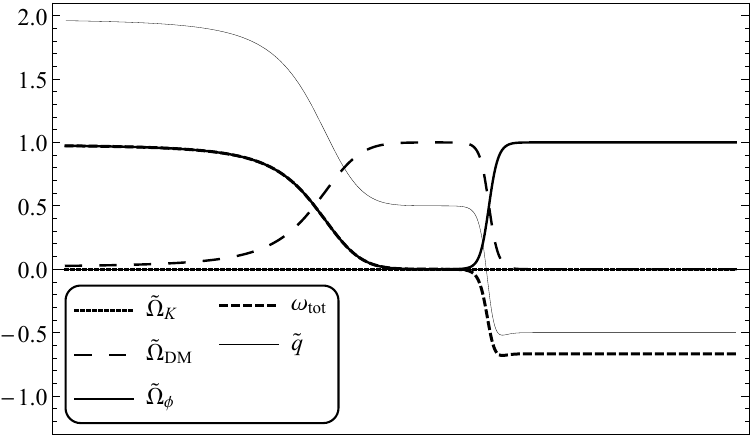}}
\subfigure[\label{fig:Evolb04}]{\includegraphics[width=0.49\linewidth]{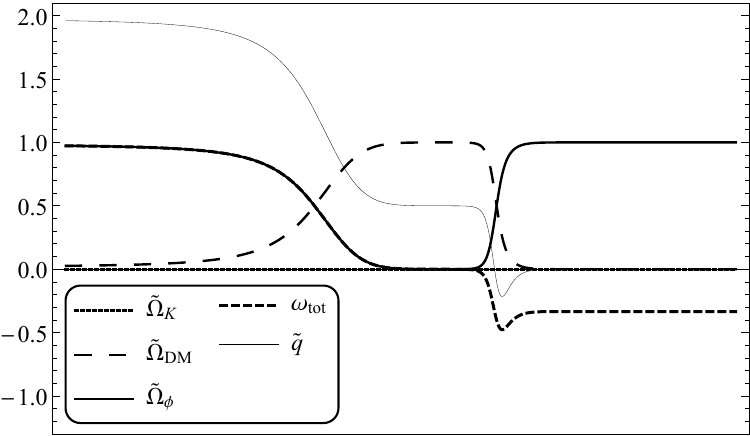}}
\end{minipage}
\caption{Evolution of the dimensionless density parameters of curvature $\tilde{\Omega}_{K}$ (dotted line), dark matter $\tilde{\Omega}_{DM}$ (large dashed line), and dark energy $\tilde{\Omega}_{\phi}$ (solid line), together with the effective equation of state $\omega_{tot}$ (dashed line), and the deceleration parameter $\tilde{q}$ (dot-dashed line) for the initial conditions $\tilde{H}(0)=1, \tilde{N}_i(0)=10^{-5}, \tilde{X}=0.99,$ and $\tilde{Y}(0)=10^{-5}$. By setting (a) $\lambda=0$, (b) $\lambda=0.5$, (c) $\lambda=1$, and (d) $\lambda=\sqrt{2}$, without the energy exchange between dark matter and dark energy, $\beta=0$.}
\label{fig:Evolb0}
\end{figure}
\noindent Figure \ref{fig:Evolb0} illustrates the evolution of the energy density parameters, effective EoS and deceleration parameter in the uncoupled limit ($\beta=0$) for four different cases, where we keep the same initial conditions and change the $\lambda$ values to $0, 0.5, 1, \sqrt{2}$. 
Across all panels, the universe passes through the same sequence of epochs. The evolution starts with kinetic dominated points $P_1/P_2$, where the field's kinetic energy is dominant component and the expansion is strongly decelerating, $\tilde q\simeq2$ and $\omega_{tot}\simeq1$.
This stiff phase is short lived as kinetic energy redshifts ( as $a^{-6}$) away faster than matter, allowing the $\tilde{\Omega}_{\text{DM}}$ to rise and establish an intermediate matter dominated era characterized by $\tilde{q} \simeq \tfrac{1}{2}$ and $\omega_{\text{tot}} \simeq 0$. As the potential term $\tilde{Y}$ grows, the scalar field takes over, driving the universe toward the scalar field dominated attractor $P_5$, where $\tilde{\Omega}_\phi \to 1$ while $\tilde{\Omega}_{\text{DM}}$ and $\tilde{\Omega}_K$ vanish. Because the initial curvature is small, $\tilde{\Omega}_K$ remains negligible throughout, keeping the evolution strictly within the isotropic subspace. With $\beta=0$ the two components do not exchange energy, and $\tilde\Omega_{DM}$ decays only because the dust dilutes as $a^{-3}$ relative to the field.

The late time acceleration is given by the fixed point $P_5$, where $\omega_{tot}=(\lambda^2-3)/3$ and $\tilde q=(\lambda^2-2)/2$. For $\lambda=0$ (panel~a) the potential is constant, the field acts as a cosmological constant, and the solution tends to de~Sitter expansion, $\omega_{tot}=-1$ and $\tilde q=-1$. Increasing the slope raises both quantities: $\lambda=0.5$ (panel b) gives $\omega_{tot}\simeq-0.92$ and $\tilde q\simeq-0.88$, and $\lambda=1$ (panel c)
gives $\omega_{tot}\simeq-0.67$ and $\tilde q=-0.5$; both still describe accelerated expansion. The case $\lambda=\sqrt2$ (panel d) is the borderline value: the attractor has $\omega_{tot}=-\tfrac13$ and $\tilde q=0$, so the expansion is exactly at the threshold between acceleration and deceleration. This is the $\lambda^2=2$ boundary identified in the stability analysis of
Section~\ref{subsec:interpretation}, beyond which $P_5$ ceases to accelerate. 

\begin{figure}[h]
\begin{minipage}[h]{1\linewidth}
\centering
\subfigure[\label{fig:Evolb05l05}]{\includegraphics[width=0.49\linewidth]{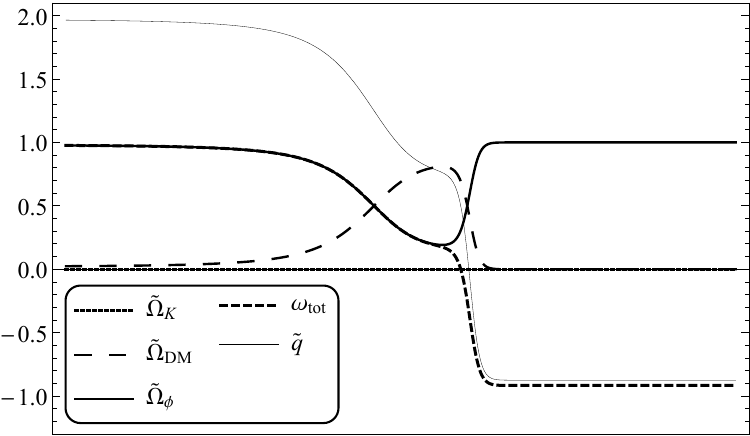}}
\subfigure[\label{fig:Odml05}]{\includegraphics[width=0.5\linewidth]{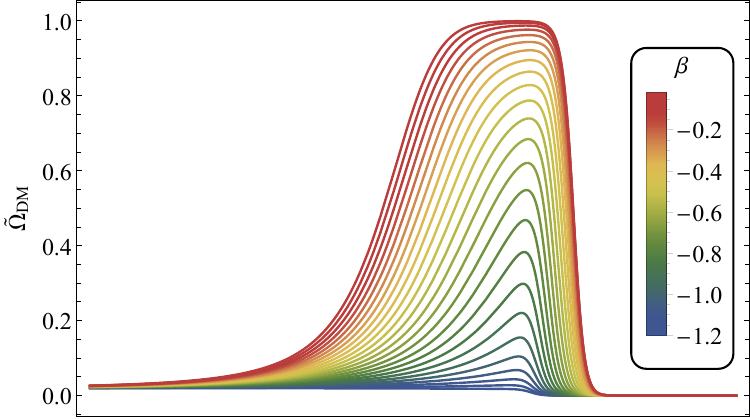}}
\end{minipage}
\caption{Evolution of the cosmological quantities for the initial conditions $\tilde{H}(0)=1, \tilde{N}_i(0)=10^{-5}, \tilde{X}=0.99,$ and $\tilde{Y}(0)=10^{-5}$ with $\lambda=0.5$. (a) Evolution of the dimensionless density parameters of curvature $\tilde{\Omega}_{K}$ (dotted line), dark matter $\tilde{\Omega}_{DM}$ (large dashed line) and dark energy $\tilde{\Omega}_{\phi}$ (solid line), together with the effective equation of state parameter $\omega_{tot}$ (dashed line) and the deceleration parameter $\tilde{q}$ (dot-dashed line) for $\beta=-0.5$. (b) Evolution of the dark matter density parameter, $\tilde{\Omega}_{DM}$, for different values of the dimensionless coupling constant $\beta$ in the range $-\sqrt{3/2}\leq\beta\leq0$. The color bar indicates the corresponding values of $\beta$.}
\label{fig:Evoll05}
\end{figure}

\noindent Figure~\ref{fig:Evoll05} introduces the dark sector coupling while keeping the same initial conditions and $\lambda=0.5$ as in the uncoupled case, so that the effect of $\beta$ can be read off directly. Panel~(a) presents a representative evolution for $\beta = -0.5$, whereas panel~(b) explores the coupling across its physical range $-\sqrt{3/2} \le \beta \le 0$. The global evolution in panel~(a) preserves the familiar three stage sequence of the uncoupled scenario: a brief kinetic dominated phase near $P_1/P_2$, an intermediate matter era where $\tilde{\Omega}_{\text{DM}}$ rises, and a subsequent approach toward the scalar field dominated attractor $P_5$ where $\tilde{\Omega}_\phi \to 1$. Because the curvature parameter $\tilde{\Omega}_K$ remains negligible throughout, the geometry stays strictly isotropic. The late time values remain identical to the uncoupled run, yielding $\omega_{\text{tot}} = (\lambda^2-3)/3 \simeq -0.92$ and $\tilde{q} = (\lambda^2-2)/2 \simeq -0.88$, because $\beta$ does not appear in the coordinates of $P_5$; the interaction modifies the transient approach toward the attractor rather than the attractor itself. However, the coupling significantly alters the matter dominated epoch. With $\beta<0$ energy is transferred from the dark matter to the scalar field, so the field acquires a non-zero velocity during matter domination and the dark-matter density parameter no longer reaches unity; instead it plateaus at the coupling-modified value $\tilde\Omega_{DM}=1-2\beta^2/3$ set by the point $P_3$, before the field takes over.

\noindent Panel~(b) makes this trend explicit by plotting $\tilde\Omega_{DM}$ for a family of couplings, with the colour bar indicating the value of $\beta$. As $|\beta|$ increases from zero, the height of the matter plateau decreases monotonically and the matter-dominated era is progressively suppressed, since a larger coupling drains more of the dark-matter energy into the field. In the limiting case $\beta=-\sqrt{3/2}$ the matter fraction at $P_3$ reaches $\tilde\Omega_{DM}=1-2\beta^2/3=0$, and the intermediate matter era disappears altogether; this is precisely the boundary $\beta^2\le3/2$ beyond which $P_3$
ceases to be physical. Across the entire range the trajectory still ends on the scalar-field attractor $P_5$, confirming that the coupling reshapes the transient matter epoch, weakening and shortening it as $|\beta|$ grows, while leaving the isotropic accelerating late-time state intact.

\begin{figure}[h]
\begin{minipage}[h]{1\linewidth}
\centering
\subfigure[\label{fig:EoSl1g1}]{\includegraphics[width=0.49\linewidth]{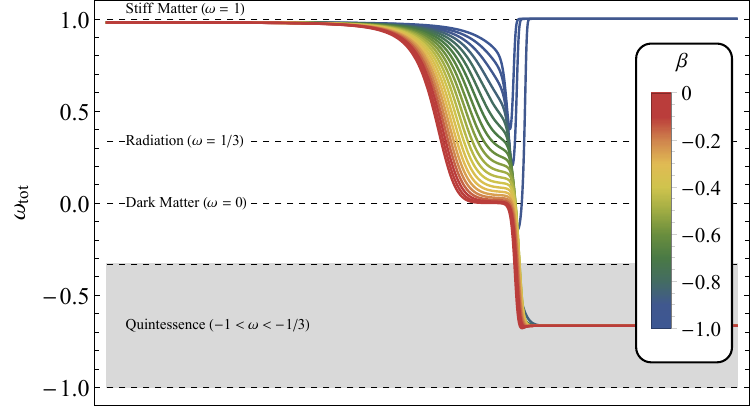}}
\subfigure[\label{fig:EoSl1g43}]{\includegraphics[width=0.5\linewidth]{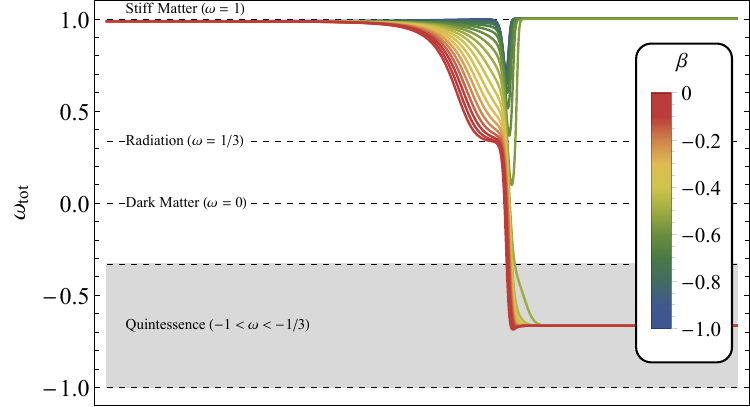}}
\end{minipage}
\caption{Evolution of the effective EoS as a function of the dimensionless coupling constant within the range $-1 \le \beta \le 0$, for initial conditions $\tilde{H}(0)=1$, $\tilde{N}_i(0)=10^{-5}$, $\tilde{X}=0.99$, and $\tilde{Y}(0)=10^{-5}$ with $\lambda=1$. Panel (a) assumes a dust background ($\gamma=1$), while panel (b) depicts a radiation background ($\gamma=4/3$). Horizontal reference lines indicate standard cosmological limits: stiff matter ($\omega=1$), radiation ($\omega=1/3$), dark matter ($\omega=0$), and dark energy ($\omega=-1$). The gray-shaded region highlights the quintessence regime ($-1 < \omega_{\text{tot}} < -1/3$), which delineates the domain of accelerated cosmic expansion. The colored curves trace the trajectories across different coupling strengths transitioning from uncoupled dynamics ($\beta=0$, red) to strong dark-sector interaction ($\beta=-1$, blue).}
\label{fig:EoS}
\end{figure}

\noindent Figure~\ref{fig:EoS} illustrates the evolution of the total effective EoS for different barotropic indices $\gamma$, examining its variation across the coupling range $-1 \le \beta \le 0$. Because the dynamical system does not include an independent variable explicitly tracking radiation energy density hence matter and radiation domination epochs cannot be isolated as separate phases. However, adjusting $\gamma$ captures the dynamical impact of both fluid contents. 

\subsection{Phase space portraits for the isotropic case}

\begin{figure}[htbp]
\begin{minipage}[h]{1\linewidth}
\centering
\subfigure[\label{fig:2Dphaseb0l0}]{\includegraphics[width=0.49\linewidth]{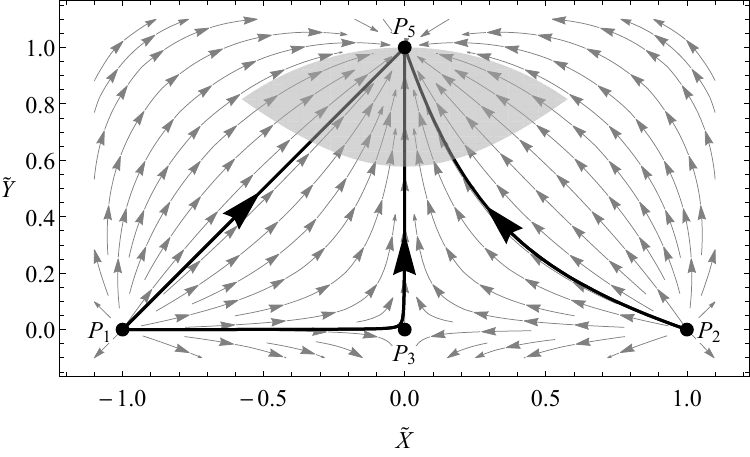}}
\subfigure[\label{fig:2Dphaseb0lsqrt2}]{\includegraphics[width=0.49\linewidth]{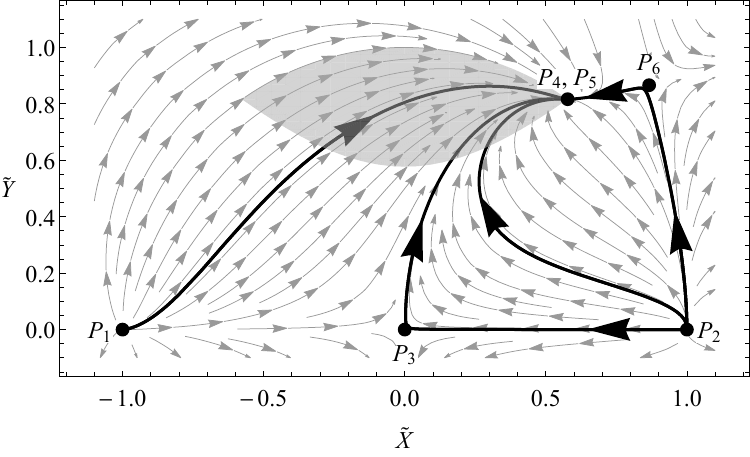}}
\end{minipage}
\caption{Phase-space trajectories in the ($\tilde{X},\tilde{Y}$) plane for $\beta=0$. The left panel (a) corresponds to $\lambda=0$, while the right panel (b) corresponds to $\lambda=\sqrt{2}$, illustrating the effect of $\lambda$ on the phase-space dynamics. The gray area denotes the region of the phase space where the universe is accelerating. Both panels are drawn onto the invariant subset $\tilde{H}=1, \tilde{\Sigma}_{\pm}=0, \tilde{N}_i=0$, where the physical region is $\tilde{X}^2+\tilde{Y}^2\leq1$. }
\label{fig:heteroclinic2Db0}
\end{figure}

\begin{figure}[htbp]
\begin{minipage}[h]{1\linewidth}
\centering
\subfigure[\label{fig:2Dphaseb05l05}]{\includegraphics[width=0.49\linewidth]{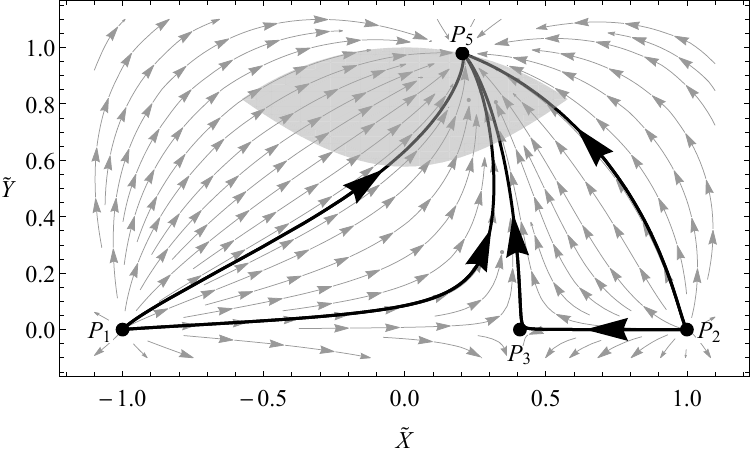}}
\subfigure[\label{fig:2Dphaseb1l05}]{\includegraphics[width=0.49\linewidth]{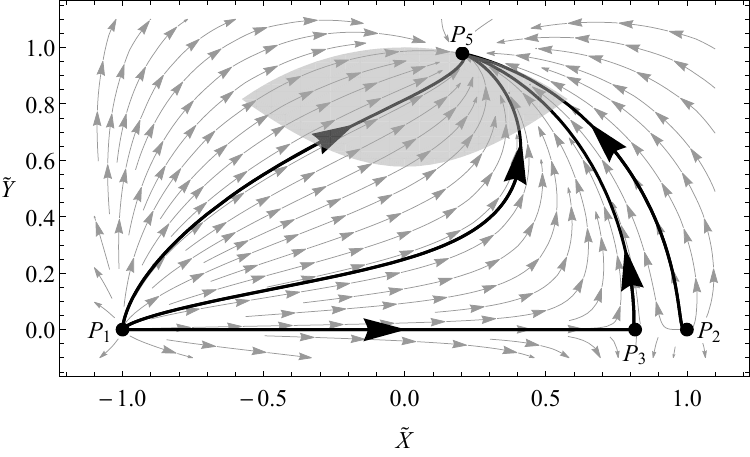}}
\end{minipage}
\caption{Phase-space trajectories in the ($\tilde{X},\tilde{Y}$) plane for $\lambda=0.5$. The left panel (a) corresponds to $\beta=-0.5$, while the right panel corresponds to $\beta=-1$. The gray area denotes the region of the phase space where the universe is accelerating. }
\label{fig:heteroclinic2D}
\end{figure}

\noindent In Figure \ref{fig:heteroclinic2Db0}, we present the phase space trajectories of the dynamical system. To establish a baseline, we first examine the uncoupled limit where DM and DE do not interact, tracking the system behavior as $\lambda$ varies from $0$ to $\sqrt{2}$. Here, the gray shaded region highlights the portion of the phase space satisfying $\omega_{\text{tot}} < -1/3$, thereby identifying the region of accelerated cosmic expansion. As $\lambda \to 0$, the fixed points $P_4$ and $P_5$ become the same point, while $P_6$ leaves the physical phase space entirely as both $\tilde{X}$ and $\tilde{Y}$ diverge to infinity. When the $\lambda = \sqrt{2}$, all fixed points appear within the phase space, but the attractor point lies precisely on the boundary of the gray shaded region since $\omega_{\text{tot}} = -1/3$.

\begin{figure}[htbp]
\begin{minipage}[h]{1\linewidth}
\centering
\subfigure[\label{fig:2Dphaseb05l15}]{\includegraphics[width=0.49\linewidth]{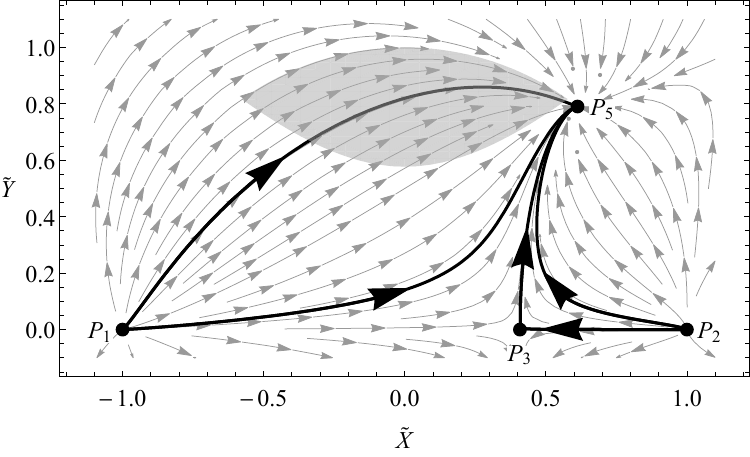}}
\subfigure[\label{fig:2Dphaseb1l15}]{\includegraphics[width=0.49\linewidth]{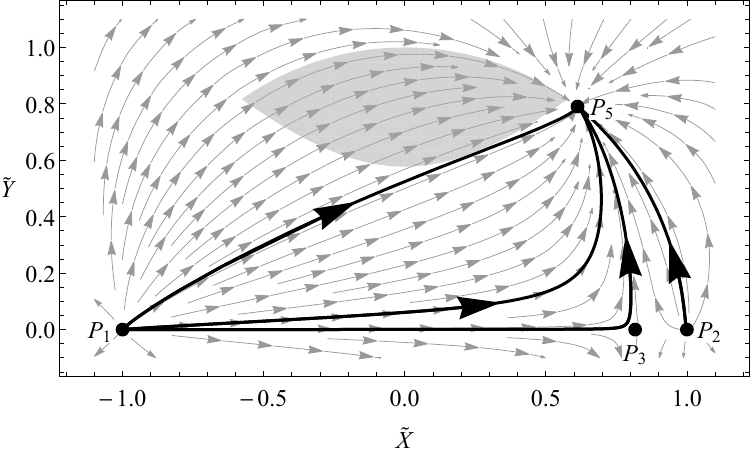}}
\end{minipage}
\caption{Phase-space trajectories in the ($\tilde{X},\tilde{Y}$) plane for $\lambda=1.5$. The left panel (a) corresponds to $\beta=-0.5$, while the right panel corresponds to $\beta=-1$. The gray area denotes the region of the phase space where the universe is accelerating.}
\label{fig:heteroclinic2D2}
\end{figure}

Figure \ref{fig:heteroclinic2D} shows the phase space trajectories for a fixed value of $\lambda(=0.5)$, illustrating the effect of varying $\beta$ from $-0.5$ to $-1$. Certain fixed points are noticeably absent from this projection due to specific geometric and parameter constraints. For instance, the curvature point $P_4$ does not appear because it carries a non zero spatial curvature $\tilde{N}_i = \sqrt{7/2}$, which forces its Hubble normalization to satisfy $\tilde{H} = \lambda/2 < 1$, placing it entirely outside the invariant boundary subset depicted in this figure. Similarly, the existence of the scaling point $P_6$ requires the parameter combination $\lambda(\beta+\lambda) \ge 3$, a condition that is violated across this range of $\beta$. In panel (a), where $\beta + \lambda = 0$, the coordinates of $P_6$ diverge to infinity, causing the point to run off the physical phase space entirely. In panel (b), where $\beta + \lambda < 0$, the corresponding expressions yield unphysical negative values ($\tilde{Y} < 0$ and $\tilde{\Omega}_{\text{DM}} < 0$), rendering the point non existent in the physical domain.

Figure~\ref{fig:heteroclinic2D2} illustrates the phase space trajectories in the $(\tilde{X}, \tilde{Y})$ plane for a steeper potential slope ($\lambda = 1.5$), here panel (a) corresponds to $\beta = -0.5$ and panel (b) corresponds to $\beta = -1$. In this steeper potential regime ($\lambda^2 = 2.25 > 2$), the closed curvature point $P_4$ fails to exist altogether, as its definition $\tilde{N}_i = \sqrt{4-2\lambda^2}$ yields an imaginary value. Additionally, the scaling point $P_6$ is entirely unphysical. Although it exists algebraically, it violates the existence requirement $\lambda(\beta+\lambda) \ge 3$. Specifically, the evaluated parameter combination gives $\lambda(\beta+\lambda) = 1.5$ in panel (a) and $0.75$ in panel (b), falling well below the required threshold and resulting in unphysical negative dark matter energy density values ($\tilde{\Omega}_{\text{DM}} = -1.5$ and $\tilde{\Omega}_{\text{DM}} = -9$, respectively).

\section{Fixed Points for the Anisotropic Case}
\label{sec:anisotropic}

We now analyze the fixed points of the full system, allowing the shear and the curvature variables to be non zero. To extract the global dynamics, we evaluate the autonomous system \eqref{eq:eom-H}--\eqref{eq:eom-Y} subject to the constraints \eqref{eq:constraintD}--\eqref{eq:friedmann}, restricting the physical phase space to $\tilde{N}_i \ge 0$ and $\tilde{Y} \ge 0$. From this point forward, we assume a pressureless dust background ($\gamma=1$) for the sake of simplicity.

Because the constraint \eqref{eq:constraintD} couples to the spatial geometry through the product $(\tilde{N}_1 \tilde{N}_2 \tilde{N}_3)^{2/3}$, the fixed points naturally divide into two structural classes. In the first class, at least one $\tilde{N}_i$ vanishes, removing the spatial curvature term from \eqref{eq:constraintD} entirely. This reduces the condition to $\tilde{H}^2=1$, meaning these fixed points are exclusively on the invariant boundary shells $\tilde{H}=\pm 1$. In the second class, all three $\tilde{N}_i > 0$; the spatial curvature remains active, allowing the dynamics to explore the interior region $|\tilde{H}| < 1$. Furthermore, similar to the isotropic case, every expanding fixed point has a contracting counterpart for $\tilde H\to-\tilde H$, since we focus on late-time acceleration we discuss only the expanding ones. 

\subsection{Physical interpretation of the fixed points}
For the anisotropic case, we obtain a broader set of fixed points. We discuss them in detail in this section, organized by the Bianchi boundary on which they reside, advancing from the spatially flat (Bianchi I) states through the intermediate single (Bianchi II) and double curvature (Bianchi VII) families, up to the fully closed Bianchi IX configuration. We have tabulated the 30 primary fixed points corresponding to the expanding case with $\tilde{N}_i \ge 0$ in Tables \ref{table:kination} and \ref{table:anisotropic}. 
\subsubsection{Kinetic dominated fixed points (Bianchi I)}
\label{subsec:kination}

For $\tilde{H}=1$, the spatially flat homogeneous sector yields 11 fixed points, labeled $P_1$ through $P_{11}$ in Table \ref{table:kination}. In these configurations, both the scalar field potential ($\tilde{Y}=0$) and the DM density ($\tilde{\Omega}_{DM}=0$) vanish entirely. Consequently, the cosmic energy budget is dominated only by the kinetic energy of the scalar field. Every fixed point in this class represents a stiff kination state characterized by $\tilde{q}=2$ and $\omega_{\text{tot}}=1$. Because all $\tilde{N}_i=0$, the Friedmann constraint \eqref{eq:friedmann} reduces precisely to the condition defining the Jacobs sphere (also known as the stiff fluid sphere \cite{WainwrightHsu1989, vandenHoogen1997}):
\begin{equation}\tilde{\Sigma}_+^2 + \tilde{\Sigma}_-^2 + \tilde{X}^2 = 1
\label{eq:kasner-sphere}
\end{equation}

\begin{table}[htbp]
\scriptsize
\centering
\begin{tabular}{|Sc|Sc|Sc|Sc|Sc|Sc|Sc|Sc|Sc|Sc|Sc|}
\hline

\hline
$\text{Point}$ & $\tilde{\Sigma}_+$ & $\tilde{\Sigma}_-$ & $\tilde{N}_1$ & $\tilde{N}_2$ & $\tilde{N}_3$ & $\tilde{X}$ & $\tilde{Y}$ & $\tilde{\Omega}_{DM}$ & $\tilde{q}$ & $\omega_{\text{tot}}$ \\ \hline

\hline
\multicolumn{11}{|l|}{\footnotesize\emph{Flat, kination phase (Bianchi~I)}}\\ \hline
$P_{1}$ & $0$ & $0$ & $0$ & $0$ & $0$ & $\pm1$ & $0$ & $0$ & $2$ & $1$ \\ \hline

$P_{2}$ & $\text{Any}$ & $0$ & $0$ & $0$ & $0$ & $\pm \sqrt{1-\tilde{\Sigma}_+^2}$ & $0$ & $0$ & $2$ & $1$ \\ \hline

$P_3$ & $\text{Any}$ & $\text{Any}$ & $0$ & $0$ & $0$ & $\pm \sqrt{1-\tilde{\Sigma}_-^2-\tilde{\Sigma}_+^2}$ & $0$ & $0$ & $2$ & $1$ \\ \hline

$P_4$ & $\text{Any}$ & $ \pm \frac{1+\tilde{\Sigma}_+}{\sqrt{3}}$ & $0$ & $0$ & $0$ & $\pm \sqrt{\frac{2}{3}} \sqrt{1-2\tilde{\Sigma}_+^2-\tilde{\Sigma}_+}$ & $0$ & $0$ & $2$ & $1$ \\ \hline

$P_{5}$ & $-1$ & $0$ & $0$ & $0$ & $0$ & $0$ & $0$ & $0$ & $2$ & $1$ \\ \hline

$P_{6}$ & $\frac{1}{2}$ & $ \pm \frac{\sqrt{3}}{2}$ & $0$ & $0$ & $0$ & $0$ & $0$ & $0$ & $2$ & $1$ \\ \hline

$P_{7}$ & $\frac{1}{2}$ & $\pm \frac{\sqrt{3-4 \tilde{X}^2}}{2} $ & $0$ & $0$ & $0$ & $\text{Any}$ & $0$ & $0$ & $2$ & $1$ \\ \hline

$P_{8}$ & $\text{Any}$ & $\pm \frac{\sqrt{-6-\lambda^2 \left(\tilde{\Sigma}_+^2-1\right)}}{\lambda}$ & $0$ & $0$ & $0$ & $\frac{\sqrt{6}}{\lambda}$ & $0$ & $0$ & $2$ & $1$ \\ \hline

\multicolumn{11}{|l|}{\footnotesize\emph{Flat, kination phase (Bianchi~I): for $\lambda^2\ge8$ }}\\ \hline
$P_{9}$ & $\frac{1}{2}$ & $\pm \frac{\sqrt{3} \sqrt{\lambda^2-8}}{2 \lambda}$ & $0$ & $0$ & $0$ & $\frac{\sqrt{6}}{\lambda}$ & $0$ & $0$ & $2$ & $1$ \\ \hline

$P_{10}$ & $-\frac{\lambda+3 \sqrt{\lambda^2-8}}{4 \lambda}$ &  $\pm \frac{\sqrt{3} \left(\lambda-\sqrt{\lambda^2-8}\right)}{4 \lambda}$ & $0$ & $0$ & $0$ & $\frac{\sqrt{6}}{\lambda}$ & $0$ & $0$ & $2$ & $1$ \\ \hline
$P_{11}$ & $\frac{3 \sqrt{\lambda^2-8}}{4 \lambda}-\frac{1}{4}$ & $\pm \frac{\sqrt{3} \left(\sqrt{\lambda^2-8}+\lambda\right)}{4 \lambda}$ & $0$ & $0$ & $0$ & $\frac{\sqrt{6}}{\lambda}$ & $0$ & $0$ & $2$ & $1$ \\ \hline

\hline
\end{tabular}
\caption{All fixed points in this table is given for $\tilde H=1$. For these fixed points, the scalar field potential and the DM energy density both vanish, $\tilde Y=0$ and $\tilde\Omega_{DM}=0$, so the dynamics is driven entirely by the field's kinetic energy; every point is consequently a stiff state with $\tilde q=2$ and $\omega_{\text{tot}}=1$, which shows the kination phase of the model. Here, entries marked with ``Any'' mark directions along which the fixed-point conditions are satisfied identically, so the fixed points are non-isolated and form continuous lines or surfaces parametrized by the free shear or curvature variable.}
\label{table:kination} 
\end{table}

In standard vacuum Bianchi models, the state space is bounded by the one-dimensional Kasner circle, defined by $\tilde{\Sigma}_+^2 + \tilde{\Sigma}_-^2 = 1$. However, since a massless scalar field dynamically mimics a stiff fluid ($\rho \propto a^{-6}$) \cite{Ibanez:1995}, the inclusion of its kinetic energy variable generalizes this boundary into a two-dimensional spherical surface. Projecting this invariant set onto the purely anisotropic shear plane yields a solid region bounded by $\tilde{\Sigma}_+^2 + \tilde{\Sigma}_-^2 \le 1$, commonly referred to in the literature as the Jacobs disc \cite{WainwrightHsu1989}. The interior of this disc contains the anisotropic scalar-field solutions, while the outer boundary represents the purely vacuum Kasner circle. Consequently, the fixed points $P_1$ through $P_8$ populate the surface of the Jacobs sphere, representing varying degrees of kinetic and shear dominance \cite{FadragasLeonSaridakis2014}. The specific points $P_9, P_{10}$, and $P_{11}$ emerge only for steep scalar field potentials satisfying $\lambda^2 \ge 8$, ensuring their shear coordinates remain strictly real. 

\noindent Crucially, none of these kinetic states serve as late time attractors. For $\tilde{H}=1$, the spatial curvature directions carry positive eigenvalues, rendering any kinetic dominated state highly unstable to the growth of spatial curvature. Their physical significance belongs strictly to the early universe: generic expanding trajectories originate from this sphere. This reflects the well known property that a massless scalar field dominates the initial singularity and successfully suppresses the chaotic Mixmaster oscillations inherent to vacuum Bianchi IX models. The dark sector coupling $\beta$ plays no role here because the DM energy density vanishes near the singularity, effectively switching off the interaction during the kination epoch.

\begin{table}[htbp]
\addtolength{\tabcolsep}{-4.5pt}
\begin{adjustwidth}{-0.5cm}{-1.5cm}
\tiny
\centering
\begin{tabular}{|Sc|Sc|Sc|Sc|Sc|Sc|Sc|Sc|Sc|Sc|Sc|}
\hline

\hline
$\text{Point}$ & $\tilde{\Sigma}_+$ & $\tilde{\Sigma}_-$ & $\tilde{N}_1$ & $\tilde{N}_2$ & $\tilde{N}_3$ & $\tilde{X}$ & $\tilde{Y}$ & $\tilde{\Omega}_{DM}$ & $\tilde{q}$ & $\omega_{\text{tot}}$ \\ \hline

\hline
\multicolumn{11}{|l|}{\footnotesize\emph{Flat, isotropic (Bianchi~I): late time acceleration}}\\ \hline
$P_{12}$ & $0$ & $0$ & $0$ & $0$ & $0$ & $\frac{\lambda}{\sqrt{6}}$ & $\frac{\sqrt{6-\lambda^2}}{\sqrt{6}}$ & $0$ & $\frac{\lambda^2-2}{2}$ & $\frac{\lambda^2-3}{3}$ \\ \hline
$P_{13}$ & $0$ & $0$ & $0$ & $0$ & $0$ & $\frac{\sqrt{\frac{3}{2}}}{\beta+\lambda}$ & $\frac{\sqrt{\beta (\beta+\lambda)+\frac{3}{2}}}{\beta+\lambda}$ & $\frac{\lambda (\beta+\lambda)-3}{(\beta+\lambda)^2}$ & $\frac{\lambda-2 \beta}{2 (\beta+\lambda)}$ & $-\frac{\beta}{\beta+\lambda}$ \\ \hline
\multicolumn{11}{|l|}{\footnotesize\emph{Single curvature (Bianchi~II) }} \\ \hline
$P_{14}$ & $\frac{\lambda-2 \beta}{8 (\beta+\lambda)}$ & $0$ & $\frac{3 \sqrt{\lambda^2-4 \beta^2}}{4 (\beta+\lambda)}$ & $0$ & $0$ & $\frac{\sqrt{\frac{3}{2}}}{\beta+\lambda}$ & $\frac{\sqrt{3} \sqrt{6 \beta^2+5 \beta \lambda+8}}{4 (\beta+\lambda)}$ & $\frac{3 \left(6 \beta \lambda+5 \lambda^2-16\right)}{16 (\beta+\lambda)^2}$ & $\frac{\lambda-2 \beta}{2 (\beta+\lambda)}$ & $-\frac{\beta}{\beta+\lambda}$ \\ \hline
$P_{15}$ & $-\frac{\lambda-2 \beta}{16 (\beta+\lambda)}$ & $\frac{\sqrt{3} (\lambda-2 \beta)}{16 (\beta+\lambda)}$ & $0$ & $0$ & $\frac{3 \sqrt{\lambda^2-4 \beta^2}}{4 (\beta+\lambda)}$ & $\frac{\sqrt{\frac{3}{2}}}{\beta+\lambda}$ & $\frac{\sqrt{3} \sqrt{6 \beta^2+5 \beta \lambda+8}}{4 (\beta+\lambda)}$ & $\frac{3 \left(6 \beta \lambda+5 \lambda^2-16\right)}{16 (\beta+\lambda)^2}$ & $\frac{\lambda-2 \beta}{2 (\beta+\lambda)}$ & $-\frac{\beta}{\beta+\lambda}$ \\ \hline
$P_{16}$ & $-\frac{\lambda-2 \beta}{16 (\beta+\lambda)}$ & $-\frac{\sqrt{3} (\lambda-2 \beta)}{16 (\beta+\lambda)}$ & $0$ & $\frac{3 \sqrt{\lambda^2-4 \beta^2}}{4 (\beta+\lambda)}$ & $0$ & $\frac{\sqrt{\frac{3}{2}}}{\beta+\lambda}$ & $\frac{\sqrt{3} \sqrt{6 \beta^2+5 \beta \lambda+8}}{4 (\beta+\lambda)}$ & $\frac{3 \left(6 \beta \lambda+5 \lambda^2-16\right)}{16 (\beta+\lambda)^2}$ & $\frac{\lambda-2 \beta}{2 (\beta+\lambda)}$ & $-\frac{\beta}{\beta+\lambda}$ \\ \hline
$P_{17}$ & $2-\frac{36}{\lambda^2+16}$ & $0$ & $\frac{6 \sqrt{(8-\lambda^2)(\lambda^2-2)}}{\lambda^2+16}$ & $0$ & $0$ & $\frac{3 \sqrt{6} \lambda}{\lambda^2+16}$ & $\frac{6 \sqrt{8-\lambda^2}}{\lambda^2+16}$ & $0$ & $8-\frac{144}{\lambda^2+16}$ & $5-\frac{96}{\lambda^2+16}$ \\ \hline
$P_{18}$ & $\frac{18}{\lambda^2+16}-1$ & $\frac{\sqrt{3} \left(\lambda^2-2\right)}{\lambda^2+16}$ & $0$ & $0$ & $\frac{6 \sqrt{(8-\lambda^2)(\lambda^2-2)}}{\lambda^2+16}$ & $\frac{3 \sqrt{6} \lambda}{\lambda^2+16}$ & $\frac{6 \sqrt{8-\lambda^2}}{\lambda^2+16}$ & $0$ & $8-\frac{144}{\lambda^2+16}$ & $5-\frac{96}{\lambda^2+16}$ \\ \hline
$P_{19}$ & $\frac{18}{\lambda^2+16}-1$ & $-\frac{\sqrt{3} \left(\lambda^2-2\right)}{\lambda^2+16}$ & $0$ & $\frac{6 \sqrt{(8-\lambda^2)(\lambda^2-2)}}{\lambda^2+16}$ & $0$ & $\frac{3 \sqrt{6} \lambda}{\lambda^2+16}$ & $\frac{6 \sqrt{8-\lambda^2}}{\lambda^2+16}$ & $0$ & $8-\frac{144}{\lambda^2+16}$ & $5-\frac{96}{\lambda^2+16}$ \\ \hline
$P_{20}$ & $2-\frac{15}{\beta^2+8}$ & $0$ & $\frac{3 \sqrt{-4 \beta^4+6 \beta^2+4}}{\beta^2+8}$ & $0$ & $0$ & $-\frac{5 \sqrt{\frac{3}{2}} \beta}{\beta^2+8}$ & $0$ & $-\frac{30 \left(\beta^2-2\right)}{\left(\beta^2+8\right)^2}$ & $8-\frac{60}{\beta^2+8}$ & $\frac{5 \beta^2}{\beta^2+8}$ \\ \hline
$P_{21}$ & $\frac{15}{2 \left(\beta^2+8\right)}-1$ & $\frac{\sqrt{3} \left(2 \beta^2+1\right)}{2 \left(\beta^2+8\right)}$ & $0$ & $0$ & $\frac{3 \sqrt{-4 \beta^4+6 \beta^2+4}}{\beta^2+8}$ & $-\frac{5 \sqrt{\frac{3}{2}} \beta}{\beta^2+8}$ & $0$ & $-\frac{30 \left(\beta^2-2\right)}{\left(\beta^2+8\right)^2}$ & $8-\frac{60}{\beta^2+8}$ & $\frac{5 \beta^2}{\beta^2+8}$ \\ \hline
$P_{22}$ & $\frac{15}{2 \left(\beta^2+8\right)}-1$ & $-\frac{\sqrt{3} \left(2 \beta^2+1\right)}{2 \left(\beta^2+8\right)}$ & $0$ & $\frac{3 \sqrt{-4 \beta^4+6 \beta^2+4}}{\beta^2+8}$ & $0$ & $-\frac{5 \sqrt{\frac{3}{2}} \beta}{\beta^2+8}$ & $0$ & $-\frac{30 \left(\beta^2-2\right)}{\left(\beta^2+8\right)^2}$ & $8-\frac{60}{\beta^2+8}$ & $\frac{5 \beta^2}{\beta^2+8}$ \\ \hline
\multicolumn{11}{|l|}{\footnotesize\emph{Double curvature (Bianchi~VII$_0$)}}\\ \hline
$P_{23}$ & $-1$ & $0$ & $0$ & $\text{Any}$ & $\text{Any}$ & $0$ & $0$ & $0$ & $2$ & $1$ \\ \hline
$P_{24}$ & $\frac{1}{2}$ & $-\frac{\sqrt{3}}{2}$ & $\text{Any}$ & $\text{Any}$ & $0$ & $0$ & $0$ & $0$ & $2$ & $1$ \\ \hline
$P_{25}$ & $\frac{1}{2}$ & $\frac{\sqrt{3}}{2}$ & $\text{Any}$ & $0$ & $\text{Any}$ & $0$ & $0$ & $0$ & $2$ & $1$ \\ \hline
$P_{26}$ & $\frac{1}{2}$ & $\frac{\sqrt{3}}{2}$ & $\frac{3 \sqrt{2 \beta^2+1}}{2 \left(\beta^2+2\right)}$ & $0$ & $\frac{3 \sqrt{2 \beta^2+1}}{2 \left(\beta^2+2\right)}$ & $0$ & $0$ & $0$ & $2$ & $1$ \\ \hline
$P_{27}$ & $\frac{1}{2}$ & $-\frac{\sqrt{3}}{2}$ & $\frac{3 \sqrt{2 \beta^2+1}}{2 \left(\beta^2+2\right)}$ & $\frac{3 \sqrt{2 \beta^2+1}}{2 \left(\beta^2+2\right)}$ & $0$ & $0$ & $0$ & $0$ & $2$ & $1$ \\ \hline
$P_{28}$ & $-1$ & $0$ & $0$ & $\frac{3 \sqrt{2 \beta^2+1}}{2 \left(\beta^2+2\right)}$ & $\frac{3 \sqrt{2 \beta^2+1}}{2 \left(\beta^2+2\right)}$ & $0$ & $0$ & $0$ & $2$ & $1$ \\ \hline
\multicolumn{11}{|l|}{\footnotesize\emph{Feature of interaction form}}\\ \hline
$P_{29}$ & $0$ & $0$ & $0$ & $0$ & $0$ & $-\sqrt{\frac{2}{3}} \beta$ & $0$ & $1-\frac{2 \beta^2}{3}$ & $\beta^2+\frac{1}{2}$ & $\frac{2 \beta^2}{3}$ \\ \hline
\multicolumn{11}{|l|}{\footnotesize\emph{Isotropic curvature (Bianchi~IX), on $\tilde H=\lambda/\sqrt2$}}\\ \hline
$P_{30}$ & $0$ & $0$ & $\sqrt{4-2\lambda^2}$ & $\sqrt{4-2\lambda^2}$ & $\sqrt{4-2\lambda^2}$ & $\frac{1}{\sqrt{3}}$ & $\sqrt{2\over3}$ & $0$ & $0$ & $-\frac{1}{3}$ \\ 
\hline
\end{tabular}
\caption{Fixed points for $\tilde H=1$ shell for which the potential, the matter, or the curvature is non-zero (points $P_{12}$--$P_{29}$), together with the point $P_{30}$ on the shell $\tilde H=\lambda/\sqrt2$. Since $\tilde H=1$ forces the curvature product $(\tilde N_1\tilde N_2\tilde N_3)^{2/3}$ to vanish, at least one $\tilde N_i=0$ on this shell.}
\label{table:anisotropic}
\end{adjustwidth}
\end{table}

\subsubsection{Isotropic late-time states ($P_{12}$ and $P_{13}$)}

The fixed points $P_{12}$ and $P_{13}$ represent the physically viable flat late-time attractors for the system. They constitute the anisotropic sector equivalents of the standard isotropic scalar field dominated and scaling points, respectively. 
The point $P_{12}$ is the uncoupled scalar field dominated state featuring $\tilde{X}=\lambda/\sqrt{6}$ and $\tilde{Y}=\sqrt{6-\lambda^2}/\sqrt{6}$. It exists provided $\lambda^2 \le 6$. Its total equation of state is $\omega_{\text{tot}}=(\lambda^2-3)/3$, guaranteeing accelerated expansion for $\lambda^2 < 2$. 

The point $P_{13}$ is the coupled scalar matter scaling point, where the energy density due to scalar field and dark matter coexist at a constant ratio. This equilibrium depends explicitly on the interaction coupling $\beta$, with the total equation of state dynamically shifted to $\omega_{\text{tot}}=-\beta/(\beta+\lambda)$ and supported by a non-zero dark matter density $\tilde{\Omega}_{DM}=[\lambda(\beta+\lambda)-3]/(\beta+\lambda)^2$. As the sole configuration where both dark components coexist at late times, it offers a direct resolution to the cosmic coincidence problem. Furthermore, it demonstrates that an interacting DM-DE model naturally enforces isotropization; assuming $P_{13}$ is a stable attractor with $\tilde{q} < 0$ (which occurs when $\lambda < 2\beta$), the interaction pulls the anisotropic universe into an accelerating phase without forcing the dark matter to dilute away completely. Finally, because both $P_{12}$ and $P_{13}$ possess zero spatial curvature and zero shear, they represent perfectly flat, isotropic universes. Any Bianchi type-IX model asymptotically approaching these fixed points has successfully damped all primordial anisotropic shear and spatial curvature.

\subsubsection{The single curvature points (Bianchi II, $P_{14}$--$P_{22}$)}

These fixed points maintain exactly one non zero structure constant, describing a universe equipped with spatial curvature along a single principal axis. Because of the inherent axis permutation symmetry of the Bianchi geometry, these points naturally appear in triplets, whice we categorise as follows:
\begin{enumerate}
    \item The points $P_{14}$, $P_{15}$ and $P_{16}$ represent anisotropic, curvature-scaling states. In this configuration, shear, spatial curvature, dark matter, and the scalar field all coexist. Notably, they share the exact same $\tilde{q}$ and $\omega_{\text{tot}}$ as the flat isotropic scaling point $P_{13}$. They represent a dynamical balance where the interacting dark sector scales proportionally with the background anisotropy and curvature. Their physical existence requires $\lambda^2 \ge 4\beta^2$ to keep the curvature variables real.

    \item For the fixed points $P_{17}$, $P_{18}$ and $P_{19}$, the energy density is due to both scalar field and curvature. It represents Bianchi type-II universe characterized by the complete depletion of dark matter ($\tilde{\Omega}_{DM}=0$). These single curvature triplets perfectly map onto the generalized Taub's orbits detailed by Fadragas, Leon, and Saridakis \cite{FadragasLeonSaridakis2014}. Specifically, point $P_{17}$ possesses zero shear along the minus-axis ($\tilde{\Sigma}_-=0$), defining it geometrically as a Locally Rotationally Symmetric (LRS) Bianchi II state. $P_{18}$ and $P_{19}$ represent the exact same physical LRS configuration, simply permuted along the remaining principal spatial axes to satisfy the threefold degeneracy of the phase space. Existing only for intermediate potential slopes restricted to $2 \le \lambda^2 \le 8$, these fixed points cannot source late-time cosmic acceleration, as their deceleration parameter $\tilde{q} = 8 - 144/(\lambda^2+16)$ remains strictly positive across this domain. Instead, their primary physical significance lies in their dynamical role within the heteroclinic network. As established in the analysis of anisotropic scalar-field cosmologies \cite{FadragasLeonSaridakis2014}, these Taub orbits form the critical invariant intermediary channels. During the early anisotropic regime, they guide the classic mixmaster dynamics, dictating the transitional bounces between the vacuum Kasner states on the Jacobs sphere before the universe eventually finds its way toward late-time isotropization.

    \item For $P_{20}$, $P_{21}$ and $P_{22}$, the scalar field potential completely vanishes ($\tilde{Y}=0$). The total energy density is due to the scalar field kinetic energy, dark matter, single axis spatial curvature and anisotropic shear. These Bianchi-II configurations describe saddles where matter and scalar field kinetic energy densities jointly dictate the shear decay and curvature dynamics.
\end{enumerate}

All fixed points within these three triplets exhibit strictly positive deceleration parameters. They do not drive cosmic acceleration but instead act as intermediate curvature supported saddle states guiding the transient evolution of the phase space trajectories.

\subsubsection{The double curvature points (Bianchi VII$_0$, $P_{23}$--$P_{28}$)}

The fixed points $P_{23}$ through $P_{28}$ belong to the double-curvature Bianchi VII$_0$ boundary, where exactly two spatial curvature variables are non-zero. For these configurations, both dark matter energy density and scalar potential energy vanish, leaving the energy budget dominated by scalar kinetic energy and spatial curvature, which dictates a stiff equation of state ($\omega_{\text{tot}}=1, \tilde{q}=2$). Due to this strong deceleration, all six double-curvature points are strictly unstable and cannot serve as late-time attractors (see Table \ref{eigenvalue_anisotropic2}). Instead, they govern the transient early-time evolution of the universe, partitioning into two distinct physical groups.

\begin{enumerate}
    \item The first set, $P_{23}$ through $P_{25}$, represents continuous, non-isolated lines of shear-curvature states where the scalar field potential completely vanishes. Because the equilibrium conditions do not constrain their active spatial curvature variables, allowing them to take any arbitrary value (denoted as ``Any''), these states form continuous channels along the phase space boundary rather than isolated points. Within the Wainwright-Hsu framework \cite{WainwrightHsu1989}, these continuous lines serve as intermediate asymptotes along the heteroclinic network, mediating transient phase-space transitions during decelerating epochs.

    \item The second set, $P_{26}$ through $P_{28}$, corresponds to isolated, Taub-like vacuum states. In Bianchi cosmology, a Taub-like configuration signifies a universe possessing Local Rotational Symmetry meaning it expands at the exact same rate in two spatial directions while evolving differently in the third. This geometric symmetry is mathematically realized here by the presence of two exactly equal active spatial curvatures. While these states reside on the same vacuum Kasner circle ($\tilde{\Sigma}_+^2 + \tilde{\Sigma}_-^2 = 1$) as the previous group, their specific curvature magnitude is not free; it is strictly regularized by the dark sector interaction coupling $\beta$. Dynamically, these Taub-like states are critically important because they act as the intermediate channels that guide the early universe's evolution. A critical dynamical feature of these points is that their eigenvalues include purely imaginary pairs (given in Table \ref{eigenvalue_anisotropic2}). Physically, these imaginary eigenvalues correspond to the rotation of the principal curvature axes. This rotation is the exact mechanism responsible for generating the chaotic Mixmaster oscillations, the continuous bouncing between successive Kasner epochs. Because they act merely as transitional ``walls'' during these early anisotropic epochs, they are highly unstable and rapidly repel the cosmic trajectory toward a different curvature regime.
\end{enumerate}

\subsubsection{The interaction point and the closed configuration ($P_{29}$ and $P_{30}$)}

The point $P_{29}$ is a flat matter kinetic scaling state characterized by $\tilde{\Omega}_{DM}=1-2\beta^2/3$ and $\omega_{\text{tot}}=2\beta^2/3$, valid for $\beta^2 \le 3/2$. It possesses zero potential and zero curvature. This state is a direct consequence of the interaction present between the DM and DE, as the scalar field kinetic energy is sustained purely by the energy drained from the dark matter sector. The fixed point $P_{30}$ is unique because it departs from the standard $\tilde{H}=1$ constraint, residing instead on the interior shell $\tilde{H}=\lambda/\sqrt{2}$. It is the only fixed point where all three spatial curvatures are non-zero and exactly equal ($\tilde{N}_i=\sqrt{4-2\lambda^2}$), representing a genuinely closed, isotropic Bianchi type-IX universe. Restricted to sufficiently flat potentials ($\lambda^2 \le 2$), this state sits exactly on the boundary between decelerated and accelerated expansion with $\tilde{q}=0$ and $\omega_{\text{tot}}=-1/3$.

Across the entire anisotropic phase space analysis, only the spatially flat points $P_{12}$ and $P_{13}$ possess the necessary stability to act as late time attractors. Every configuration having anisotropic shear or spatial curvature ultimately proves unstable, ensuring that these features remain strictly transient. The stiff kination states govern the initial singularity, the single and double curvature families provide intermediate saddle routes, and the fully closed Bianchi IX point sits marginally at the acceleration threshold. The inevitable end state for a generic expanding trajectory is thus an isotropic, scalar field dominated or scalar matter scaling universe, confirming that the coupled Bianchi IX model generically isotropizes and achieves late time acceleration.

\subsection{Evolution Plots}

In this subsection, we present the evolution plots for the full anisotropic system to examine the individual and combined impacts of initial anisotropic shear ($\tilde{\Sigma}^2$), spatial curvature and dark sector interactions on cosmic expansion and late time isotropization.

\subsubsection{Effect of Shear}
\begin{figure}[htbp]
\begin{minipage}[h]{1\linewidth}
\centering
\subfigure[\label{}]{\includegraphics[width=0.49\linewidth]{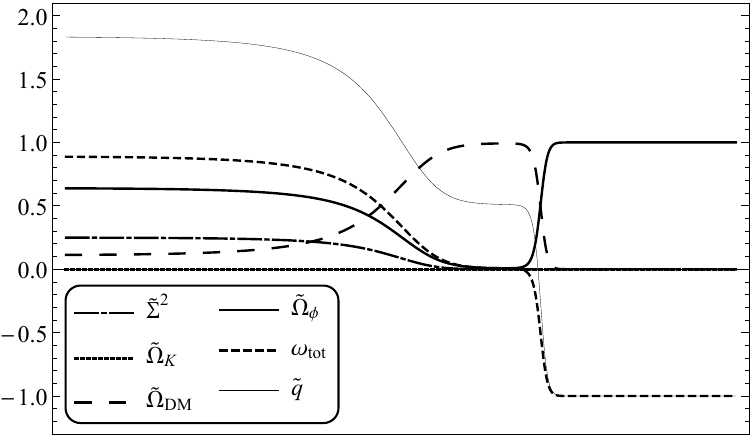}}
\subfigure[\label{}]{\includegraphics[width=0.49\linewidth]{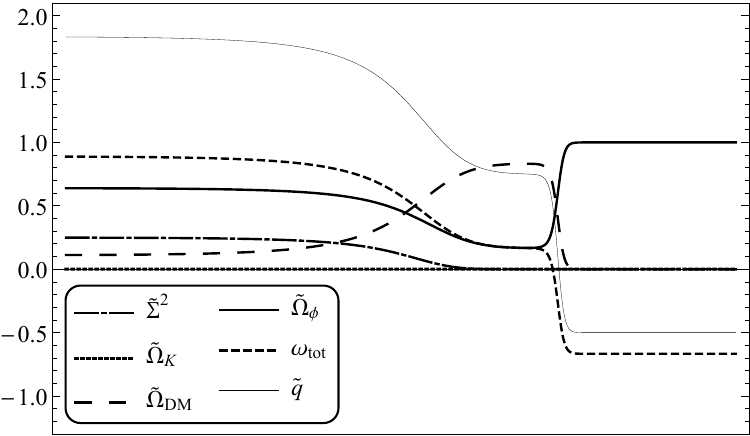}}
\subfigure[\label{}]{\includegraphics[width=0.49\linewidth]{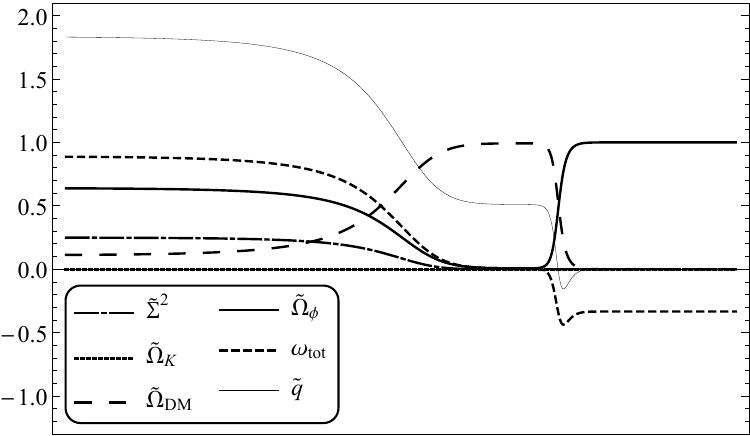}}
\subfigure[\label{}]{\includegraphics[width=0.49\linewidth]{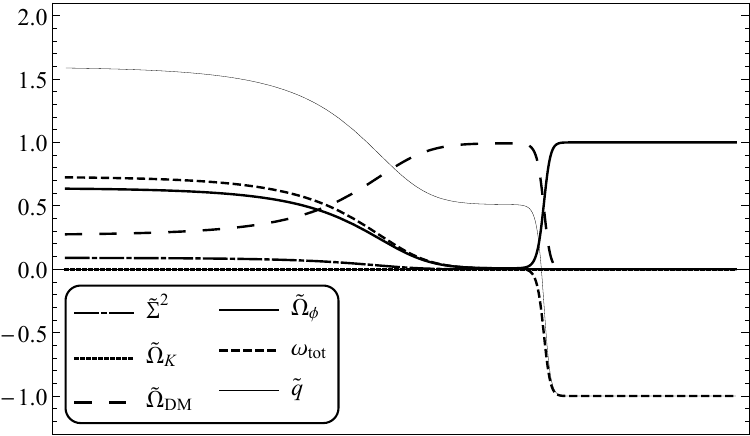}}
\subfigure[\label{}]{\includegraphics[width=0.49\linewidth]{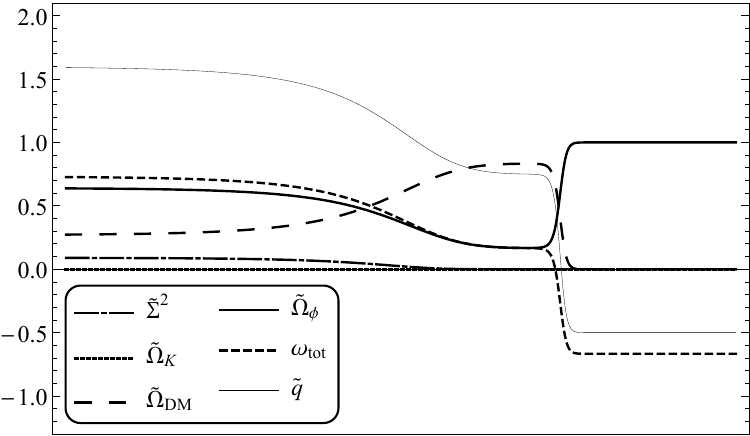}}
\subfigure[\label{}]
{\includegraphics[width=0.49\linewidth]{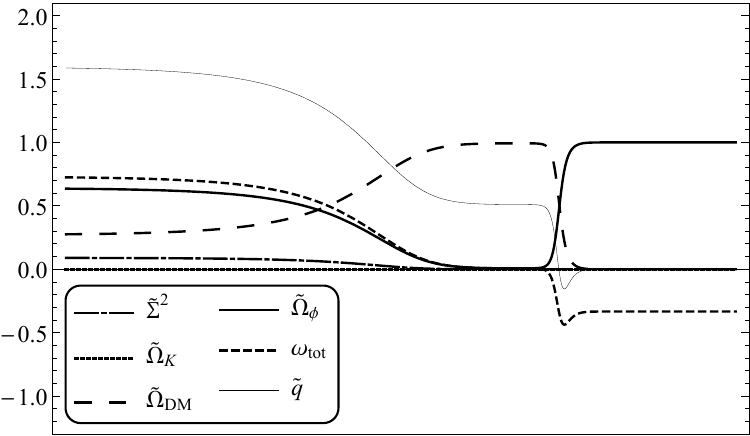}}
\end{minipage}
\caption{Evolution of the cosmological parameters illustrating the effect of initial anisotropic shear. All panels share the background initial conditions $\tilde{H}(0)=\sqrt{1-\frac{1}{4}(\tilde{N}_1 \tilde{N}_2 \tilde{N}_3)^{2/3}}$, $\tilde{N}_i(0)=10^{-5}$, $\tilde{X}(0)=0.8$, and $\tilde{Y}(0)=10^{-5}$. The top panels showcase evolution dominated by the plus-component of shear ($\tilde{\Sigma}_{+}(0)=0.5, \tilde{\Sigma}_{-}(0)=10^{-5}$) for (a) $\beta=-0.1, \lambda=0$, (b) $\beta=-0.5, \lambda=1$, and (c) $\beta=-0.1, \lambda=\sqrt{2}$. The bottom panels showcase evolution dominated by the minus-component of shear ($\tilde{\Sigma}_{+}(0)=10^{-5}, \tilde{\Sigma}_{-}(0)=0.3$) for (d) $\beta=-0.1, \lambda=0$, (e) $\beta=-0.1, \lambda=\sqrt{2}$, and (f) $\beta=-0.5, \lambda=1$. In all cases, the early universe undergoes a transient, shear-driven stiff phase ($\tilde{q} \simeq 2$, $\omega_{\text{tot}} \simeq 1$). For a flat potential ($\lambda=0$; panels a and d), the system smoothly recovers standard $\Lambda$CDM-like behavior at late times. As the potential slope $\lambda$ increases, the transition to the accelerating epoch is delayed and the final accelerating strength is reduced. Regardless of the initial shear distribution, the scalar field ultimately washes out the anisotropies, guaranteeing late-time isotropization.}
\label{shearpm}
\end{figure}

\begin{figure}[htbp]
\begin{minipage}[h]{1\linewidth}
\centering
\subfigure[\label{}]{\includegraphics[width=0.49\linewidth]{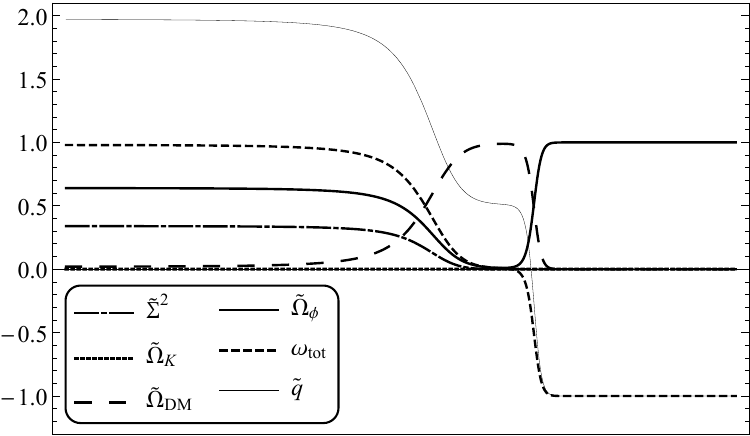}}
\subfigure[\label{}]{\includegraphics[width=0.49\linewidth]{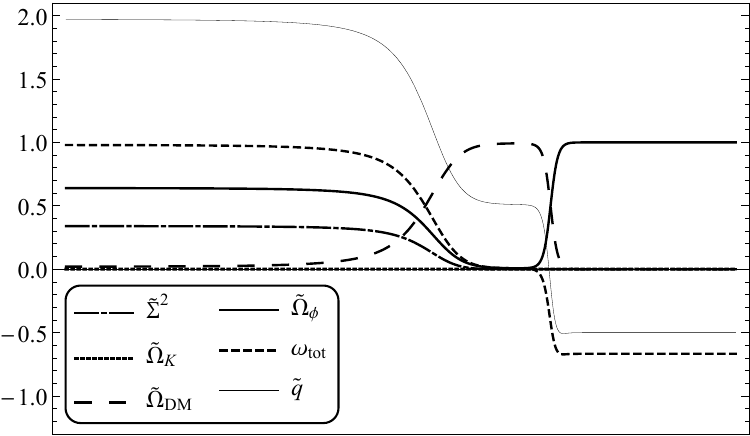}}
\subfigure[\label{}]{\includegraphics[width=0.49\linewidth]{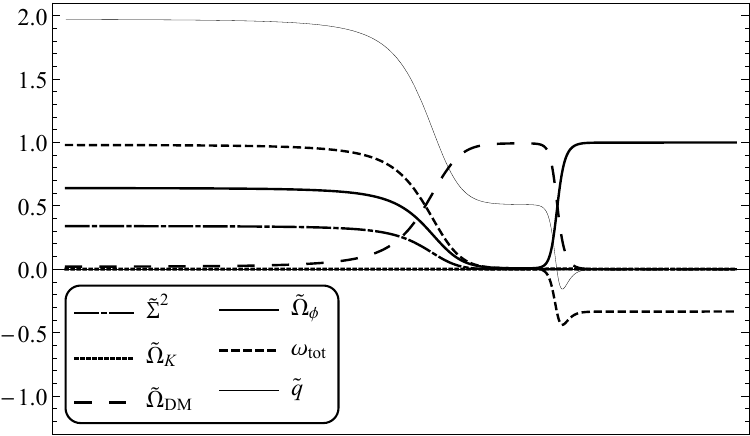}}
\subfigure[\label{}]{\includegraphics[width=0.49\linewidth]{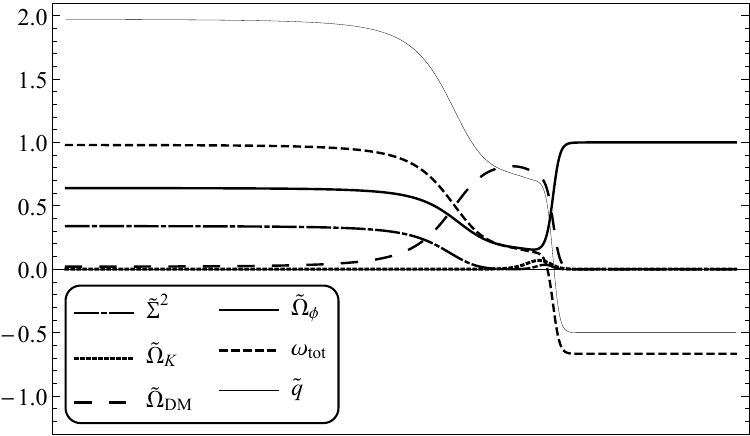}}
\end{minipage}
\caption{Evolution of the cosmological parameters under the combined influence of both plus and minus initial shear components. The shared background initial conditions are $\tilde{H}(0)=\sqrt{1-\frac{1}{4}(\tilde{N}_1 \tilde{N}_2 \tilde{N}_3)^{2/3}}$, $\tilde{\Sigma}_{+}(0)=0.5$, $\tilde{\Sigma}_{-}(0)=0.3$, $\tilde{N}_i(0)=10^{-5}$, $\tilde{X}(0)=0.8$, and $\tilde{Y}(0)=10^{-5}$. The panels correspond to (a) $\beta=-0.1, \lambda=0$, (b) $\beta=-0.1, \lambda=1$, (c) $\beta=-0.1, \lambda=\sqrt{2}$, and (d) $\beta=-0.5, \lambda=1$. The simultaneous presence of both active shear components drives a strong initial anisotropic stiff epoch ($\tilde{q} \simeq 2$, $\omega_{\text{tot}} \simeq 1$). Similar to the isolated shear cases, a flat potential (panel a) successfully mimics standard $\Lambda$CDM late-time behavior. Increasing the potential slope $\lambda$ (panels b and c) diminishes the accelerating strength and delays the transition into the scalar-field-dominated phase. A stronger interaction coupling $\beta$ (panel d) shifts the intermediate matter scaling dynamics, yet the scalar field ultimately drives the system to a fully isotropic ($\tilde{\Sigma}_{\pm} \to 0$), accelerating late-time attractor across all parameter choices.}
\label{shearall}
\end{figure}

Figures \ref{shearpm} and \ref{shearall} illustrate the evolution of cosmological parameters when non-zero initial anisotropic shear is introduced alongside interaction within the dark sector. In the early universe, the anisotropic shear acts geometrically like a stiff fluid, dominating the energy budget and driving strong deceleration with $\tilde{q} \simeq 2$ and $\omega_{\text{tot}} \simeq 1$. Because shear redshifts away faster than matter ($\rho \propto a^{-6}$), this anisotropic stiff phase is transient. As the shear dilutes, the dark matter density parameter rises to establish an intermediate matter-dominated era. Eventually, the scalar field potential takes over, damping the shear and driving the system toward the isotropic, scalar field dominated late-time attractor. 

In Figure \ref{shearpm}, we present the numerical evolution of the cosmological parameters, establishing a consistent line convention that is maintained across all anisotropic evolution plots in this work. The solid black line represents the dark energy density ($\tilde{\Omega}_\phi$), which clearly captures the initial kination epoch at early times before eventually growing to drive the late-time accelerating phase.

The dot-dashed line tracks the shear contribution ($\tilde{\Sigma}^2$), highlighting that the early universe is highly anisotropic before this shear rapidly redshifts away. The loosely dashed line traces the fractional dark matter density ($\tilde\Omega_{DM}$), while the finely dashed line indicates the spatial curvature density ($\tilde\Omega_K$). Finally, the thin solid line plots the deceleration parameter ($\tilde{q}$), allowing us to pinpoint exactly when the universe crosses the threshold from early deceleration into late-time acceleration.

\subsubsection{Effect of $\tilde{N}_i$}

\begin{figure}[htbp]
\begin{minipage}[h]{1\linewidth}
\centering
\subfigure[\label{}]{\includegraphics[width=0.49\linewidth]{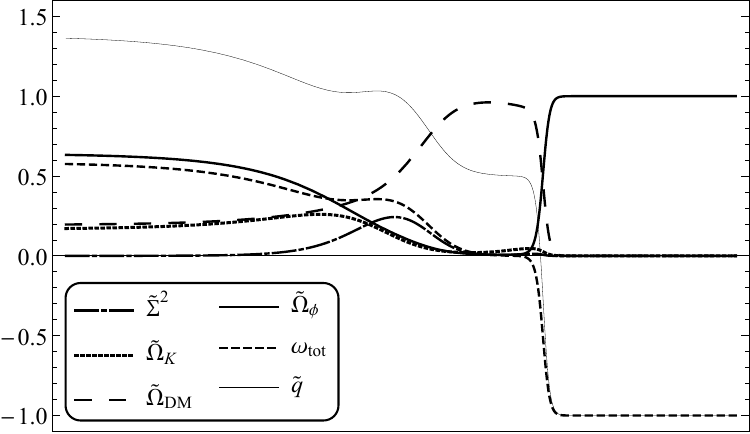}}
\subfigure[\label{}]{\includegraphics[width=0.49\linewidth]{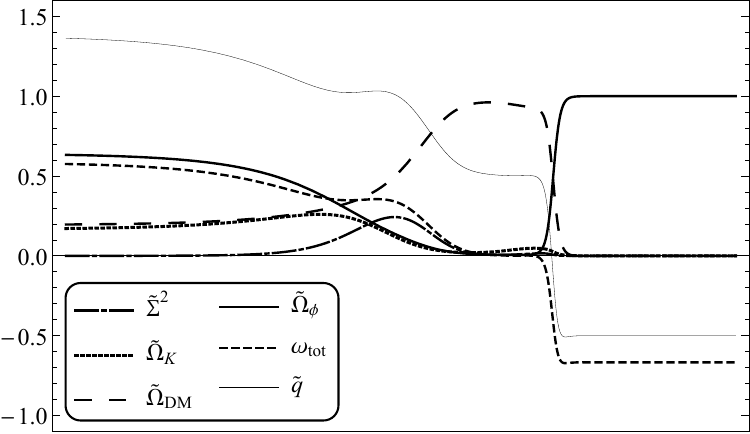}}
\subfigure[\label{}]{\includegraphics[width=0.49\linewidth]{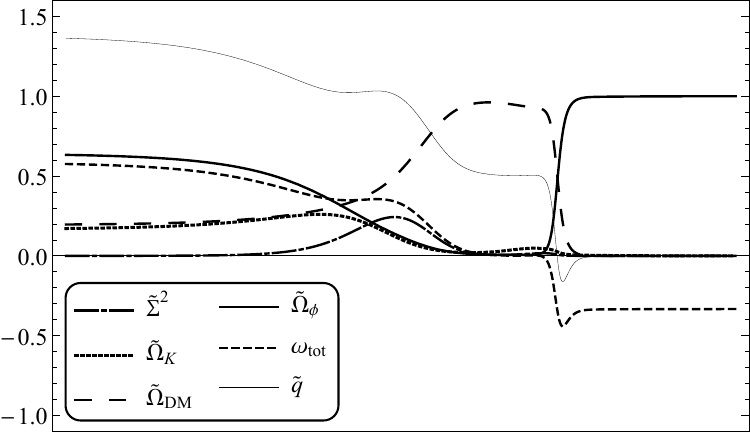}}
\subfigure[\label{}]{\includegraphics[width=0.49\linewidth]{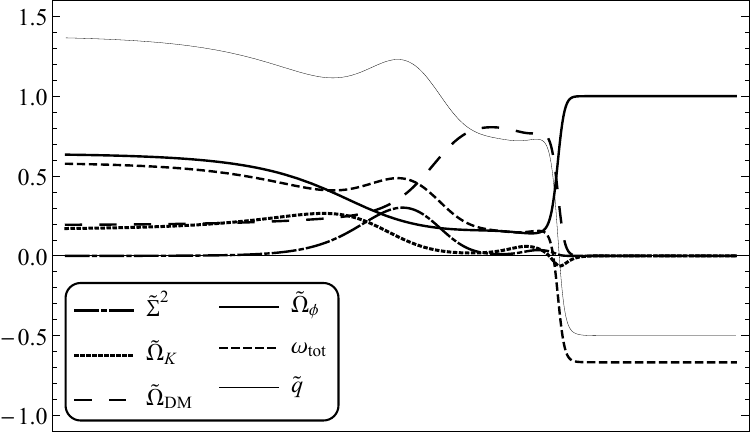}}
\end{minipage}
\caption{Evolution of the cosmological parameters illustrating the effect of a single dominant spatial curvature component. The shared initial conditions are $\tilde{H}(0)=\sqrt{1-\frac{1}{4}(\tilde{N}_1 \tilde{N}_2 \tilde{N}_3)^{2/3}}$, $\tilde{\Sigma}_{+}(0)=10^{-5}$, $\tilde{\Sigma}_{-}(0)=10^{-5}$, $\tilde{N}_1(0)=\sqrt{2}$, $\tilde{N}_2(0)=10^{-5}$, $\tilde{N}_3(0)=10^{-5}$, $\tilde{X}(0)=0.8$, and $\tilde{Y}(0)=10^{-5}$. The panels display variations in the dark sector parameters: (a) $\beta=-0.1, \lambda=0$, (b) $\beta=-0.1, \lambda=1$, (c) $\beta=-0.1, \lambda=\sqrt{2}$, and (d) $\beta=-0.5, \lambda=1$. Despite the large initial spatial curvature along one principal axis, the scalar field successfully overcomes the curvature, driving the universe to an isotropic late-time accelerating state.}
\label{fig:curvature_single}
\end{figure}

\begin{figure}[H]
\begin{minipage}[h]{1\linewidth}
\centering
\subfigure[\label{}]{\includegraphics[width=0.49\linewidth]{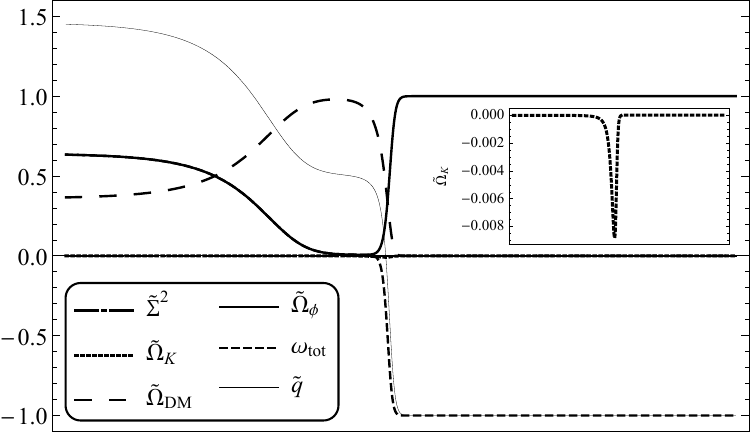}}
\subfigure[\label{}]{\includegraphics[width=0.49\linewidth]{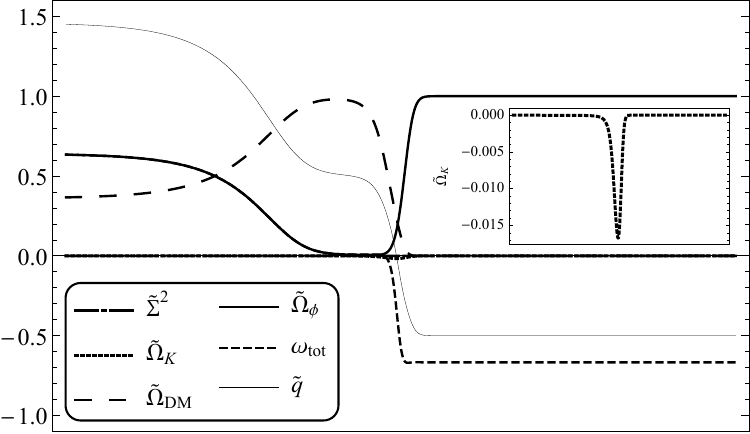}}
\end{minipage}
\caption{Evolution of the cosmological parameters under the influence of two active spatial curvature components. The initial conditions are set to $\tilde{H}(0)=\sqrt{1-\frac{1}{4}(\tilde{N}_1 \tilde{N}_2 \tilde{N}_3)^{2/3}}$, $\tilde{\Sigma}_{+}(0)=10^{-5}$, $\tilde{\Sigma}_{-}(0)=10^{-5}$, $\tilde{N}_1(0)=\sqrt{2}$, $\tilde{N}_2(0)=\sqrt{2}$, $\tilde{N}_3(0)=10^{-5}$, $\tilde{X}(0)=0.8$, and $\tilde{Y}(0)=10^{-5}$. The panels correspond to a fixed coupling $\beta=-0.1$ with varying potential slopes: (a) $\lambda=0$ and (b) $\lambda=1$. The presence of two large curvature modes alters the early intermediate dynamics, but the trajectory still successfully converges to the isotropic scalar-field attractor.}
\label{fig:curvature_double}
\end{figure}

\begin{figure}[H]
\begin{minipage}[h]{1\linewidth}
\centering
\subfigure[\label{}]{\includegraphics[width=0.49\linewidth]{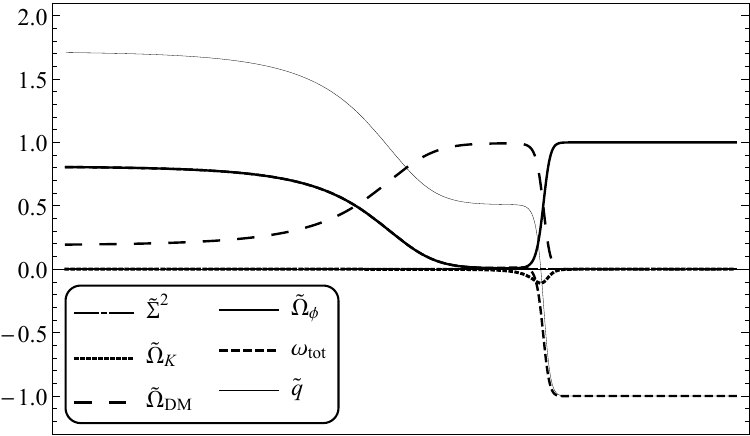}}
\subfigure[\label{}]{\includegraphics[width=0.49\linewidth]{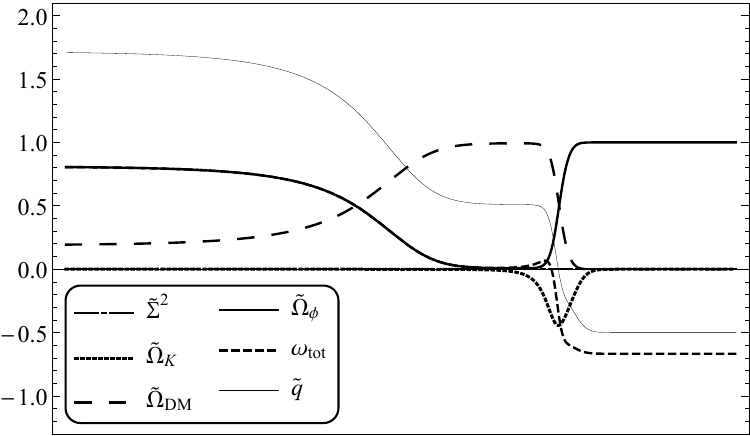}}
\end{minipage}
\caption{Evolution of the cosmological parameters for a genuinely closed geometry where all three spatial curvatures are active. The initial conditions are $\tilde{H}(0)=\sqrt{1-\frac{1}{4}(\tilde{N}_1 \tilde{N}_2 \tilde{N}_3)^{2/3}}$, $\tilde{\Sigma}_{+}(0)=10^{-5}$, $\tilde{\Sigma}_{-}(0)=10^{-5}$, $\tilde{N}_1(0)=\tilde{N}_2(0)=\tilde{N}_3(0)=0.01$, $\tilde{X}(0)=0.9$, and $\tilde{Y}(0)=10^{-5}$. The panels show a coupling of $\beta=-0.1$ with (a) $\lambda=0$ and (b) $\lambda=1$. The fully closed curvature acts uniformly but remains subdominant to the quintessence field, which smoothly initiates accelerated expansion without triggering a recollapse.}
\label{fig:curvature_triple}
\end{figure}

Figures \ref{fig:curvature_single}, \ref{fig:curvature_double} and \ref{fig:curvature_triple} display the dynamical evolution of the cosmological parameters when non-zero spatial curvature variables are active. By activating single, double and fully isotropic spatial curvatures, these plots systematically demonstrate how the closed geometric structure of the Bianchi IX influences the cosmic trajectory. 

During the early epochs, the spatial curvature actively contributes to the Friedmann constraint ($\tilde{\Omega}_K > 0$), acting alongside the scalar field's kinetic energy and any residual anisotropic shear. As seen in Figures \ref{fig:curvature_single} and \ref{fig:curvature_double}, introducing a large initial curvature along one or two principal axes noticeably changes the intermediate cosmological transitions. The presence of this geometric curvature alters the transient balance between the kinetic and matter components, slightly reshaping the onset and plateau of the matter dominated era compared to the purely spatially flat cases.  
However, despite this intermediate influence, the spatial curvature remains dynamically subdominant in the long term for expanding trajectories governed by sufficiently flat potentials ($\lambda^2 < 2$). As the universe expands, the energy density associated with the spatial curvature redshifts. The scalar field potential eventually overtakes all other components, initiating a phase of dark energy domination ($\hat{\Omega}_\phi \to 1$). This transition is marked by the deceleration parameter dropping below zero ($\tilde{q} < 0$), signaling the onset of late time cosmic acceleration. 

The most physically significant feature demonstrated across all three configurations is the successful avoidance of the recollapse. In standard GR without dark energy, the positive spatial curvature inherent to closed topologies inevitably halts expansion and forces a turnaround. In this interacting model, the quintessence field successfully violates the strong energy condition, driving exponential expansion before the spatial curvature can reach the critical threshold required to force a recollapse. This accelerated expansion rapidly flattens the spatial geometry, driving $\tilde{\Omega}_K \to 0$ and completely damping the geometric anisotropies. Consequently, whether the curvature is isolated to a single axis or acts uniformly across all three (Figure \ref{fig:curvature_triple}), the universe successfully isotropizes and converges strictly to the flat, accelerating scalar field late-time attractor.

\subsubsection{Effect of Combined Shear and Spatial Curvature}

\begin{figure}[htbp]
\begin{minipage}[h]{1\linewidth}
\centering
\subfigure[\label{}]{\includegraphics[width=0.49\linewidth]{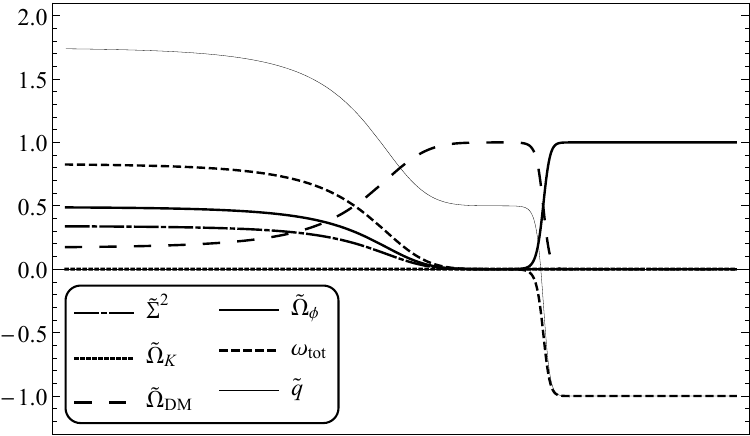}}
\subfigure[\label{}]{\includegraphics[width=0.49\linewidth]{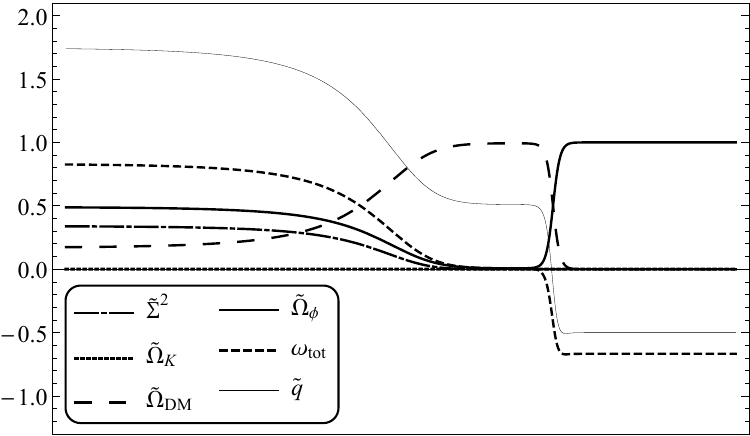}}
\subfigure[\label{}]{\includegraphics[width=0.49\linewidth]{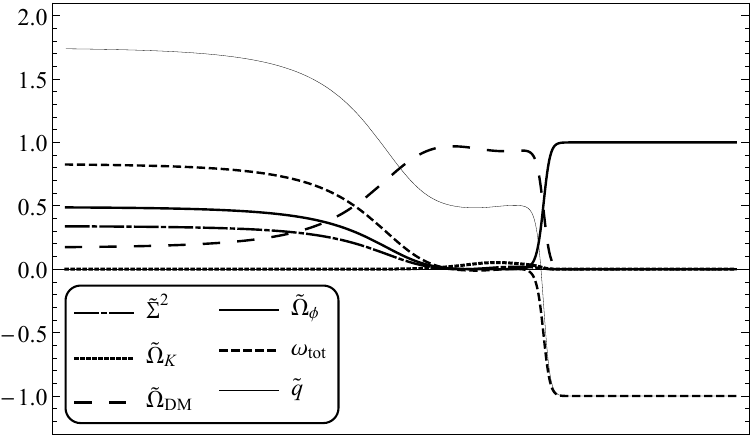}}
\subfigure[\label{}]{\includegraphics[width=0.49\linewidth]{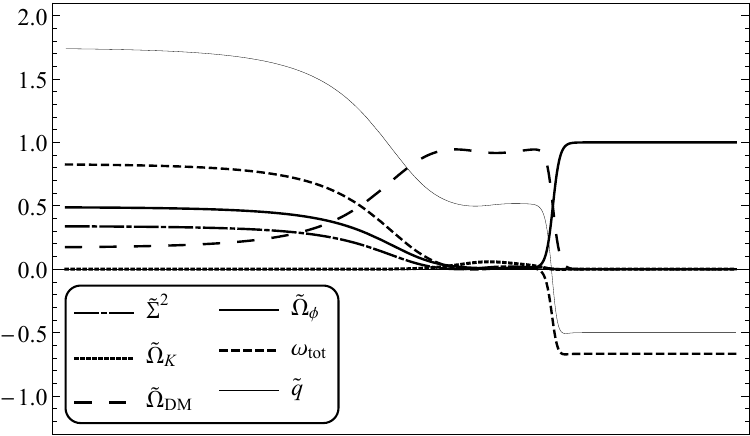}}
\subfigure[\label{}]{\includegraphics[width=0.49\linewidth]{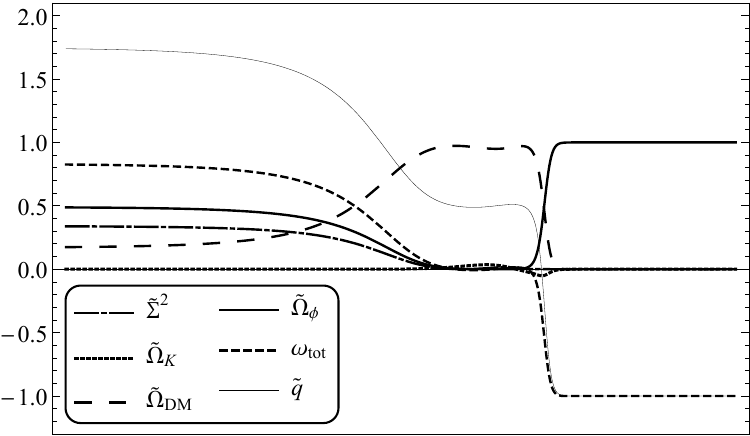}}
\subfigure[\label{}]{\includegraphics[width=0.49\linewidth]{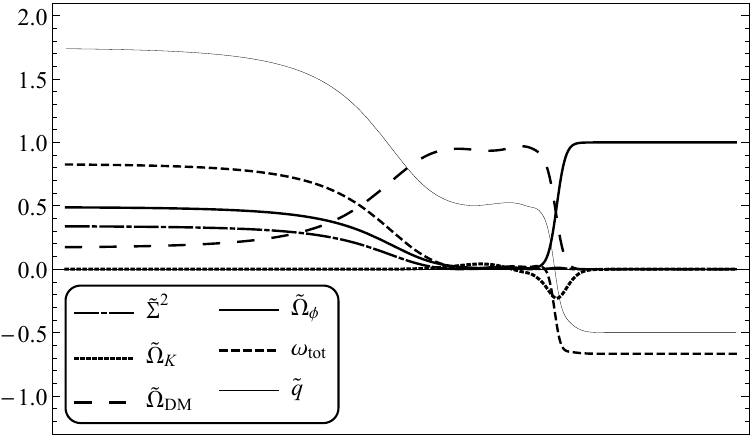}}
\end{minipage}
\caption{Evolution of the cosmological parameters under the combined influence of both initial anisotropic shear and spatial curvature. The shared initial conditions are $\tilde{H}(0) = \sqrt{1 - \frac{1}{4}(\tilde{N}_1\tilde{N}_2\tilde{N}_3)^{2/3}}$, $\tilde{\Sigma}_+(0)=0.5$, $\tilde{\Sigma}_-(0)=0.3$, $\tilde{X}(0)=0.7$, and $\tilde{Y}(0)=10^{-5}$. The left panels (a, c, e) correspond to an uncoupled scalar field with a constant potential ($\beta=0, \lambda=0$), mimicking standard $\Lambda$CDM late-time behavior. The right panels (b, d, f) correspond to an interacting dark sector with a steeper potential ($\beta=-0.1, \lambda=1$). The rows vary the active spatial curvature components: (a, b) single active curvature with $\tilde{N}_1(0)=0.01$ and $\tilde{N}_{2}(0)=\tilde{N}_{3}(0)=10^{-5}$; (c, d) double active curvature with $\tilde{N}_{1}(0)=\tilde{N}_{2}(0)=0.01$ and $\tilde{N}_3(0)=10^{-5}$; (e, f) fully closed isotropic curvature with $\tilde{N}_{1}(0)=\tilde{N}_{2}(0)=\tilde{N}_{3}(0)=0.01$. In all cases, the scalar field overcomes the initial geometric complexities, ensuring the shear and curvature completely decay and driving the universe to a fully isotropic, accelerating late-time attractor.}
\label{fig:combined_shear_curvature}
\end{figure}

Figure \ref{fig:combined_shear_curvature} presents the full dynamical evolution when both initial anisotropic shear and non-zero spatial curvature variables are simultaneously active. By combining these effects, we test the ultimate robustness of the cosmic no-hair theorem for our dynamical system. 
In all scenarios, the early universe is heavily dominated by the anisotropic shear, which acts geometrically as a stiff fluid and drives a strong initial deceleration phase. Despite the simultaneous presence of positive spatial curvature, this curvature remains subdominant and does not recollapse for small values of $\beta$.

As the universe expands, the shear components rapidly redshift away, allowing the dark matter density to momentarily rise and establish an intermediate plateau. The left panels (a, c, e) demonstrate that for an uncoupled system with a flat potential ($\beta=0, \lambda=0$), the dynamics perfectly recover standard $\Lambda$CDM behavior, smoothly transitioning into a de Sitter-like late-time state ($\omega_{\text{tot}} \to -1$). The right panels (b, d, f) show that introducing an interaction ($\beta=-0.1$) and a steeper potential ($\lambda=1$) alters the intermediate matter scaling and reduces the final accelerating strength, but does not disrupt the overall stability of the attractor.   

Crucially, as we move down the rows, progressing from single curvature (a, b), to double curvature (c, d), to fully closed isotropic curvature (e, f), the qualitative evolution remains identical. The repulsive dynamics driven by the quintessence field successfully overcome both the initial anisotropic perturbations and the closed geometry. The shear and curvature components completely decay, confirming that the interacting Bianchi IX model generically isotropizes and converges to a stable, accelerating late-time attractor.   

\subsubsection{Recollapse for Strong Dark Sector Couplings}

\begin{figure}[htbp]
    \centering
    \includegraphics[width=0.6\linewidth]{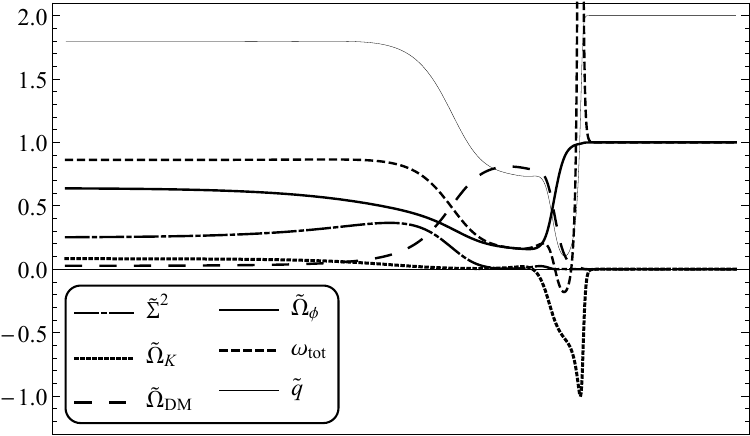}
    \caption{Evolution of the cosmological parameters illustrating a cosmic recollapse induced by a strong dark sector interaction. The initial conditions are $\tilde{H}(0)=\sqrt{1-\frac{1}{4}(\tilde{N}_1 \tilde{N}_2 \tilde{N}_3)^{2/3}}$, $\tilde{\Sigma}_{+}(0)=0.5$, $\tilde{\Sigma}_{-}(0)=10^{-5}$, $\tilde{N}_1(0)=1$, $\tilde{N}_2(0)=10^{-5}$, $\tilde{N}_3(0)=10^{-5}$, $\tilde{X}(0)=0.8$, and $\tilde{Y}(0)=10^{-5}$, evaluated for $\beta=-0.5$ and $\lambda=1$. Because the strong coupling sustains the kinetic energy of the scalar field, the total effective equation of state fails to violate the strong energy condition early enough. Consequently, the positive spatial curvature ($\tilde{\Omega}_K$) overtakes the background expansion, leading to a cosmic turnaround ($H=0$) and subsequent recollapse, marked by the sharp divergence of the dynamical variables.}    \label{fig:recollapse}
\end{figure}

While sufficiently flat potentials generally allow the quintessence field to drive late-time acceleration and evade the closed-universe recollapse conjecture, this isotropization mechanism can break down if the dark sector interaction is too strong. Figure \ref{fig:recollapse} illustrates this behavior for a large coupling magnitude ($= -0.5$) combined with a steep potential ($\lambda=1$) and an active spatial curvature along a single principal axis. 

Physically, a large negative coupling rapidly transfers energy from the dark matter fluid into the scalar field. Crucially, this energy injection sustains the scalar field's kinetic energy and prevents the potential energy from dominating the cosmic energy budget early enough. Because the kinetic term acts dynamically like a stiff fluid, the total EoS  fails to properly drop below the critical $-1/3$ threshold required to violate the strong energy condition. Without sustained accelerated expansion to dilute the geometric anisotropies, the positive spatial curvature eventually overtakes the energy budget. This forces the expansion rate to zero ($H=0$) at the moment of maximum expansion, causing the trajectories of the dynamical variables to sharply break as the universe inevitably recollapses into a contracting phase.

\section{Comparison Table with existing literature} \label{sec:comparison}

\begin{table}[H]
\renewcommand{\arraystretch}{2}
\addtolength{\tabcolsep}{-2.1pt}
\tiny
\begin{tabular}{cccccc}
\hline
S. No. & Model & Spatial Curvature & Matter Content & Late-Time Attractors & Ref \\ \hline
1. & \makecell{LRS Bianchi I and \\
LRS Bianchi III} & For $k=0$ and $k=-1$ & Scalar field + barotropic fluid & \makecell{Isotropized de Sitter } & \cite{FadragasLeonSaridakis2014} \\

\vspace{0.3cm}

2. & LRS Bianchi III & For $k=-1$ & \makecell{ Scalar field \\ (generalized harmonic potential) \\ + barotropic fluid} & \makecell{Matter-dominated flat FLRW ($0\le\gamma\le\tfrac{2}{3}$); \\ matter--curvature scaling ($\tfrac{2}{3}<\gamma<1$); \\ Bianchi III flat spacetime ($1\le\gamma\le2$)} & \cite{Leon:2021a} \\

\vspace{0.3cm}

3. & Bianchi VII${}_h$ + Scalar Field & For $k=-1$ & Uncoupled Quintessence & Marginally Inflationary FRW & \cite{vandenHoogen1997} \\

\vspace{0.3cm}

4. & \makecell{Kantowski--Sachs} & For $k=1$ & \makecell{Uncoupled scalar field \\ (generalized harmonic potential) \\ + barotropic fluid} & \makecell{Contracting non-flat LRS Kasner and \\ Taub ($0\le\gamma\le2$); flat matter-dominated \\ FLRW ($0\le\gamma\le\tfrac{2}{3}$)} & \cite{Leon:2021c} \\

\vspace{0.3cm}

5. & LRS Bianchi V & For $k=-1$ & \makecell{Uncoupled scalar field \\ (generalized harmonic potential) \\ + barotropic fluid} & \makecell{Geometry and $\gamma$-dependent: \\ matter-dominated / curvature-dominated \\ (Milne-type) attractors} & \cite{MillanoLeon:2024} \\

\vspace{0.3cm}

6. & LRS Bianchi I & For $k=0$ & \makecell{Coupled scalar field, \\ $Q=Q(H,\dot\phi,\rho_{\rm dm},\rho_{\rm de})$ \\ (harmonic potential)} & \makecell{Parameter-dependent accelerated \\ attractors (averaged system)} & \cite{MillanoPaliathanasis:2025} \\

\vspace{0.3cm}

7. & LRS Bianchi I & For $k=0$ & Coupled Quintessence & \makecell{de Sitter} & \cite{RathoreSingh:2023} \\

\vspace{0.3cm}

8. & \makecell{LRS Bianchi IX and \\
Kantowski-Sachs} & For $k=1$ & \makecell{Perfect fluid} & \makecell{Expansion from singularity \\ followed by recollapse} & \cite{Heinzle:2005} \\

\vspace{0.3cm}

9. & Bianchi IX (non-tilted) & For $k= 1$ & \makecell{Viscous fluid ($p=w\mu$, \\
constant bulk and shear viscosity) } & \makecell{ Flat FLRW; Bianchi type II solution; \\
Jacobs disc; Bianchi type $\text{VII}_0$ solutions} & \cite{Kohli:2014} \\

\vspace{0.3cm}

10. & \makecell{Bianchi class A \\ (types I--IX)} & \makecell{$k=0,\pm1$ \\ (type-dependent)} & Perfect fluid & \makecell{Equilibrium set of Kasner, \\ Bianchi~II, Collins--Ellis and \\ isotropic FLRW points} & \cite{WainwrightHsu:1989} \\

\vspace{0.3cm}

11. & \makecell{Bianchi models \\ (class A, incl.\ IX)} & \makecell{$k=0,\pm1$} & \makecell{Uncoupled scalar field \\ (exponential potential)} & \makecell{Isotropic power-law inflationary \\ FLRW ($\lambda^2<2$); anisotropic \\ Feinstein--Ib\'a\~nez attractor ($\lambda^2>2$)} & \cite{Ibanez:1995} \\

\vspace{0.3cm}

12. & LRS Bianchi I & For $k=0$ & \makecell{Coupled scalar field \\ (non-canonical, interacting \\ dark sector) + dust} & \makecell{Quintessence scalar-field dominated \\ solution; accelerated scaling \\ attractor} & \cite{Bhanja:2023} \\

\vspace{0.3cm}

13. & Bianchi IX & For $k= 1$ & Uncoupled Quintessence & \makecell{flat FLRW (for $\lambda^2<2$)} & \cite{vdHoogenOlasagasti1999} \\

\vspace{0.3cm}

14. & LRS Bianchi I & For $k=0$ & \makecell{Interacting dark sector \\ (non linear coupling \\ $Q \propto \rho_{\rm dm}\rho_{\rm de}$)} & \makecell{Stable accelerating \\ scaling solutions} & \cite{SharifSaleem2012} \\

\vspace{0.3cm}

15. & LRS Bianchi I & For $k=1$ & \makecell{Modified $f(R,T)$ gravity \\ + perfect fluid} & \makecell{Isotropic power law \\ FLRW universes} & \cite{MoraesSahoo2017} \\

\vspace{0.3cm}

16. & \textbf{This Work} & \textbf{For $k=1$} & \textbf{Coupled Quintessence} & \makecell{\textbf{Isotropic scalar field dominated universe;} \\ \textbf{coupled scalar matter scaling universe}} & --- \\[1.5ex]

\hline

\hline
\end{tabular}
\caption{Comparison of the present work with representative dynamical-systems studies of anisotropic cosmologies. For each model we list the geometry, the sign of the spatial curvature $k$, the matter content, and the late-time attractors. }
\label{table:comparison}
\end{table}

Table~\ref{table:comparison} places the present work within the dynamical-systems literature on spatially homogeneous cosmologies. The existing literature is largely differentiated on the basis of the spatial geometry (flat $k=0$, open $k=-1$, or closed $k=1$), the degree of anisotropy modeled and the existence of a dark sector coupling. To the best of our knowledge, no previous study has occupied the corner of this parameter space in which all three nontrivial choices are made together: a closed Bianchi IX geometry, the full anisotropic sector and an interacting dark sector. This is perhaps understandable as the positive spatial curvature causes these models to recollapse and the standard expansion-normalized variables break down at maximum expansion, which has tended to keep closed models separate from the coupled-quintessence studies carried out in flat geometry. The present analysis addresses precisely this gap.

\noindent In the two limiting cases the dynamical system reproduces known results. First where by setting $\beta=0$ together with $\tilde\Omega_{\rm DM}=0$ recovers the uncoupled Bianchi~IX system \cite{vdHoogenOlasagasti1999}. And second in the flat, shear-free limit $\tilde N_i=0$, $\tilde\Sigma_\pm=0$, $\tilde H=1$, the anisotropy and curvature drop out and the equations collapse to the coupled quintessence system of Amendola~\cite{Amendola2000}, in which dark matter and dark energy interact on a spatially flat FLRW background. Recovering both limits confirms that our coupled, anisotropic model is a consistent generalization of these two established systems.

\section{Conclusion}
\label{sec:conclusions}

We have presented a comprehensive fixed-point analysis of a coupled Bianchi IX cosmology, evaluating the local stability of all equilibrium states to map the global dynamics and asymptotic behavior of the universe in the presence of an interacting dark sector. The full eight-dimensional phase space reveals a rich dynamical landscape, encompassing spatially flat FLRW solutions ($\tilde{N}_i=0$, $\tilde{\Sigma}_\pm=0$), Bianchi type-II states (where one spatial curvature variable is active), Bianchi type-$\text{VII}_0$ boundary configurations (characterized by two active curvature variables), Bianchi IX states (where all curvature variables are active) and shear dominated Kasner-like kinetic spheres ($\tilde{N}_i=0$), spanning both expanding and contracting regimes. From this global dynamical framework, several fundamental insights into the nature of late-time cosmic evolution emerge.

First, by analyzing closed spatial curvature within the more general Bianchi IX framework, generalizes isotropic $k=+1$ FLRW models by allowing spatial curvature to vary independently across all three principal directions alongside anisotropic shear degrees of freedom over $S^3$ topology, we establish that positive curvature does not universally enforce a cosmic recollapse. While ordinary pressureless matter in GR causes rapid deceleration that allows spatial curvature to dominate and trigger a Big Crunch turnaround, a quintessence field with a sufficiently flat potential ($\lambda^2 < 2$) provides a robust mechanism to violate the strong energy condition ($\omega_{\text{tot}} < -1/3$). The resulting accelerated expansion exponentially dilutes both spatial curvature and shear, drawing generic trajectories away from recollapse and toward isotropization. Crucially, however, our global mapping reveals that this mechanism is sensitive to the interaction strength: a sufficiently strong coupling ($\beta \ll 0$) sustains scalar field kinetic energy, delaying strong energy condition violation and restoring spatial curvature driven turnaround ($H=0$) and recollapse.

Second, the dark sector interaction fundamentally reshapes the intermediate matter-dominated epoch and the timeline of isotropization. By continuously transferring energy from dark matter to dark energy, the coupling forces the scalar field to maintain a non-zero velocity during matter domination. While stronger interactions suppress the peak fractional dark matter density, they simultaneously accelerate the onset of dark energy domination. Furthermore, although the anisotropic coupled scaling state ($P_{13}$) acts as a multidimensional saddle rather than a final late-time attractor, the transient scaling regime it supports offers a dynamical framework for mitigating the cosmic coincidence problem without requiring extreme fine-tuning of initial conditions.

Finally, regarding asymptotic phase-space trajectories, generic expanding models emerge from an early, shear dominated Kasner state, traverse a complex network of single and double curvature saddle points and ultimately converge onto a spatially flat, accelerating FLRW attractor. This complete phase space mapping demonstrates that the cosmic no-hair theorem extends successfully to closed, anisotropic topologies in the presence of a scalar-mediated dark sector interaction, proving that a highly anisotropic closed universe generically evolves into the smooth, accelerating cosmos observed today.

\FloatBarrier
\acknowledgments
The authors would like to express sincere gratitude to Dr. Kinjal Banerjee, Dr. Chakrit Pongkitivanichkul and Dr. Peera Simakachorn for insightful discussions and valuable feedback during the preparation of this work. This work has received scholarship under the Post Doctoral Training for Frontier Research from Khon Kaen University, Thailand (Grant No. PD2569-18).

\raggedbottom
\renewcommand{\cleardoublepage}
{\clearpage}
\appendix
\section{Appendix 1}
\label{appendix1}

\subsection{Eigenvalues of the isotropic sector}

\begin{table}[H]
\addtolength{\tabcolsep}{1pt}
\begin{adjustwidth}{-1cm}{-1cm}
    \centering
    \begin{tabular}{|Sc|Sc|Sc|}
    \hline

    \hline
    Point & Eigenvalues & Stability \\ \hline
    
    \hline
    $P_{1}$ & $4,2,2,2,0,0,3-\sqrt{6} \beta ,\sqrt{\frac{3}{2}} \lambda +3$ & Unstable \\ \hline

    $P_{2}$ & $4,2,2,2,0,0,\sqrt{6} \beta +3,3-\sqrt{\frac{3}{2}} \lambda$ & \makecell{Unstable for $\lambda\leq\sqrt{6},\,\beta\geq-\sqrt{3\over2}$ \\ Saddle for $\lambda>\sqrt{6},\,\beta<-\sqrt{3\over2}$} \\ \hline
    
    $P_{3}$ & \makecell{$\beta ^2-\frac{3}{2},\beta ^2-\frac{3}{2},\beta ^2-\frac{3}{2},\beta ^2+\frac{1}{2},\beta ^2+\frac{1}{2},$ \\ 
    $\beta^2+\frac{1}{2},2 \beta ^2+1,\beta  (\beta +\lambda )+\frac{3}{2}$} &  Saddle \\ \hline
    $P_{4}$ & \makecell{$0,\frac{2 \beta -\lambda }{\sqrt{2}},-\frac{\sqrt{8-3 \lambda ^2}+\lambda }{\sqrt{2}},\frac{\sqrt{8-3 \lambda^2}-\lambda }{\sqrt{2}},-\frac{\sqrt{9 \lambda ^2-16}+\lambda }{\sqrt{2}},-\frac{\sqrt{9 \lambda ^2-16}+\lambda}{\sqrt{2}},$\\
    $\frac{\sqrt{9 \lambda ^2-16}-\lambda }{\sqrt{2}},\frac{\sqrt{9 \lambda ^2-16}-\lambda}{\sqrt{2}}$} & Saddle \\ \hline
    $P_{5}$ & \small{\makecell{$\frac{1}{2} \left(\lambda ^2-6\right),\frac{1}{2} \left(\lambda ^2-6\right),\frac{1}{2} \left(\lambda^2-6\right),\frac{1}{2} \left(\lambda ^2-2\right),\frac{1}{2} \left(\lambda ^2-2\right),$\\
    $\frac{1}{2} \left(\lambda^2-2\right),\lambda ^2-2,\lambda  (\beta +\lambda )-3$}} & Stable \\ \hline
    $P_{6}$ & \makecell{$1-\frac{3 \beta }{\beta +\lambda },\frac{\lambda -2 \beta }{2 (\beta +\lambda )},\frac{\lambda -2 \beta }{2
   (\beta +\lambda )},\frac{\lambda -2 \beta }{2 (\beta +\lambda )},-\frac{3 (2 \beta +\lambda )}{2 (\beta +\lambda
   )},-\frac{3 (2 \beta +\lambda )}{2 (\beta +\lambda )},$ \\ $\frac{\pm\sqrt{-3 \left(32 \beta ^2+21\right) \lambda ^2+12
   \beta  \left(9-4 \beta ^2\right) \lambda +36 \left(5 \beta ^2+6\right)-48 \beta  \lambda ^3}-6 \beta -3 \lambda }{4
   (\beta +\lambda )}$} & Saddle \\  \hline
    
   \hline
    \end{tabular}
    \caption{Eigenvalues of the fixed points for the isotropic case for the expanding universe evaluated for $\gamma=1$ and the resulting stability classification in the physical regime $\beta<0$, $\lambda^2<2$.} 
    \label{table:2}
\end{adjustwidth}
\end{table}

\subsection{Stability of the anisotropic fixed points}
\label{subsec:aniso-stability}

To determine the stability of the system, we linearize the autonomous equations around each fixed point and evaluate the eigenvalues of the Jacobian matrix. The constraint \eqref{eq:friedmann} is used to eliminate one variable, so each fixed point carries eight eigenvalues. The results are listed in Tables~\ref{eigenvalue_anisotropic1} and~\ref{eigenvalue_anisotropic2} for $\lambda\ge0$ and $\beta\le0$, and stability is stated only within the domain of existence of each point.

\begin{table}[H]
\tiny
\addtolength{\tabcolsep}{-2.5pt}
\begin{adjustwidth}{-2cm}{-2cm}
\centering
\begin{tabular}{|Sc|Sc|Sc|}
\hline
Point & Eigenvalues & Stability \\

\hline
$P_{1}$ & $4,2,2,2,0,0,3\pm\sqrt{6}\beta,3\mp\sqrt{\frac{3}{2}} \lambda$ & \makecell{Unstable for $0\le\lambda<\sqrt6$, $-\sqrt{\frac32}<\beta\le0$ }\\ 

\hline
$P_{2}$ & $4,0,0,2(1+\tilde{\Sigma}_+),2(1+\tilde{\Sigma}_+),2-4\tilde{\Sigma}_+,3\pm\beta\sqrt{6-6\tilde{\Sigma}_+^2},3\mp\sqrt{\frac{3}{2}}\lambda\sqrt{1-\tilde{\Sigma}_+^2}$ & \makecell{$\left(\tilde{X}=\sqrt{1-\tilde{\Sigma}_{+}^2}\right)$ \\ Source for $-1<\tilde{\Sigma}_{+}<\frac{1}{2}$, $0\le\lambda<\frac{\sqrt6}{\sqrt{1-\tilde{\Sigma}_{+}^2}}$ \\ and $-\frac{\sqrt{3/2}}{\sqrt{1-\tilde{\Sigma}_{+}^2}}<\beta\leq0$ \\ $\left(\tilde{X}=-\sqrt{1-\tilde{\Sigma}_{+}^2}\right)$ \\ Source for $-1<\tilde{\Sigma}_{+}<\frac{1}{2}$; saddle for $\tilde{\Sigma}_{+}>\frac{1}{2}$ \\ for all $\lambda\ge0$, $\beta\le0$} \\

\hline
$P_{3}$ & \makecell{$4,0,0,2+2\tilde{\Sigma}_{+}-2\sqrt{3}\tilde{\Sigma}_{-},2(1+\tilde{\Sigma}_{+}+\sqrt{3}\tilde{\Sigma}_{-}),2-4\tilde{\Sigma}_{+},$\\$3\pm\sqrt{6}\beta\sqrt{1-\tilde{\Sigma}_{+}^2-\tilde{\Sigma}_{-}^2},3\mp\sqrt{3\over2}\lambda\sqrt{1-\tilde{\Sigma}_{+}^2-\tilde{\Sigma}_{-}^2}$} & \makecell{$(\tilde{X}=\sqrt{1-\tilde{\Sigma}_{+}^2-\tilde{\Sigma}_{-}^2})$ \\ Source for $\tilde{\Sigma}_{+}<\frac{1}{2}$, $1+\tilde{\Sigma}_{+}\pm\sqrt{3}\tilde{\Sigma}_{-}>0$ \\ and $0\leq\lambda<\frac{\sqrt{6}}{\tilde X}$, $-\frac{\sqrt{3/2}}{\tilde X}<\beta\leq0$ \\ $(\tilde{X}=-\sqrt{1-\tilde{\Sigma}_{+}^2-\tilde{\Sigma}_{-}^2})$ \\ Source for $\tilde{\Sigma}_{+}<\frac{1}{2}$, $1+\tilde{\Sigma}_{+}\pm\sqrt{3}\tilde{\Sigma}_{-}>0$} \\

\hline
$P_{4}$ & $4,0,0,0,4(1+\tilde{\Sigma}_{+}),2-4\tilde{\Sigma}_{+},3\mp2\beta\sqrt{1-\tilde{\Sigma}_{+}-2\tilde{\Sigma}_{+}^2},3\pm\lambda\sqrt{1-\tilde{\Sigma}_{+}-2\tilde{\Sigma}_{+}^2}$ & \makecell{Non-hyperbolic. For $-1<\tilde{\Sigma}_{+}<\frac{1}{2}$: \\ $(\tilde X<0)$: non-zero eigenvalues positive for all $\lambda\geq0,\beta\leq0$ \\ $(\tilde X>0)$: non-zero eigenvalues positive for \\ $\lambda\sqrt{1-\tilde{\Sigma}_{+}-2\tilde{\Sigma}_{+}^2}<3$, $2\beta\sqrt{1-\tilde{\Sigma}_{+}-2\tilde{\Sigma}_{+}^2}>-3$} \\

\hline
\makecell{$P_{5}$ \\ $P_{6}$} & $6,4,3,3,0,0,0,0$ & Non-hyperbolic, unstable \\

\hline
$P_{7}$ & $4,0,0,0,3-\sqrt{9-12\tilde{X}^2},3+\sqrt{9-12\tilde{X}^2},3+\sqrt{6}\beta\tilde{X},3-\sqrt{3\over2}\lambda\tilde{X}$ & \makecell{Non-hyperbolic. Non-zero eigenvalues positive for \\ $-\frac{\sqrt{3}}{2}\leq\tilde{X}<0$, all $\lambda\geq0$, $\beta\leq0$, \\ or $0<\tilde{X}\leq\frac{\sqrt{3}}{2}$, $0\leq\lambda<\frac{\sqrt{6}}{\tilde{X}}$, $-\frac{\sqrt{3/2}}{\tilde{X}}<\beta\leq0$} \\ 

\hline
$P_{8}$ & $0,0,0,4,3+\frac{6\beta}{\lambda},2-4\tilde{\Sigma}_{+},\frac{2\left(\lambda+\lambda\tilde{\Sigma}_{+}\mp\sqrt{3\lambda^2(1-\tilde{\Sigma}_{+}^2)-18}\right)}{\lambda}$ & \makecell{Non-hyperbolic. Non-zero eigenvalues positive for \\ $-1<\tilde{\Sigma}_{+}<\frac{1}{2}$, $-\frac{\lambda}{2}<\beta\leq0$, \\ $\frac{\sqrt{6}}{\sqrt{1-\tilde{\Sigma}_{+}^2}}\leq\lambda<\frac{3}{\sqrt{1-\tilde{\Sigma}_{+}-2\tilde{\Sigma}_{+}^2}}$ }\\ 

\hline
\makecell{$P_{9}$ \\ $P_{10}$ \\ $P_{11}$} & $0,0,0,0,4,3+\frac{6\beta}{\lambda},3\pm\frac{3\sqrt{\lambda^2-8}}{\lambda}$ & \makecell{Non-hyperbolic. Non-zero eigenvalues positive \\ for $\lambda\geq2\sqrt{2}$, $-\frac{\lambda}{2}<\beta\leq0$ }\\ 

\hline
\end{tabular}
\caption{Eigenvalues and stability of the fixed points $P_1$--$P_{11}$ of Table~\ref{table:kination}, all of which lie on the Jacobs sphere \eqref{eq:kasner-sphere} with $\tilde H=1$ and $\tilde Y=\tilde\Omega_{DM}=\tilde N_i=0$. Results are given for $\lambda\ge0$ and $\beta\le0$. Where a row carries $\pm$ or $\mp$, the upper sign corresponds to $\tilde X>0$ and the lower sign to $\tilde X<0$.}
\label{eigenvalue_anisotropic1}
\end{adjustwidth}
\end{table}

\paragraph{Jacobs sphere ($P_1$--$P_{11}$):} 
At any point on the kinetic-dominated sphere \eqref{eq:kasner-sphere}, the eight eigenvalues are
\begin{equation}
\left\{0,\; 0,\; 4,\; 2-4\tilde\Sigma_+,\; 2\left(1+\tilde\Sigma_+\pm\sqrt{3}\,\tilde\Sigma_-\right),\; 3-\sqrt{\tfrac{3}{2}}\lambda\tilde{X},\; 3+\sqrt{6}\beta\tilde{X}\right\}.
\label{eq:jacobs-eigs}
\end{equation}
These fixed points corresponds to unstable states. The coupling $\beta$ affects only one eigenvalue ($3+\sqrt{6}\beta\tilde{X}$), shrinking the source region on the $\tilde{X}>0$ hemisphere while leaving the $\tilde{X}<0$ side unaffected.

\paragraph{Flat isotropic points ($P_{12}, P_{13}, P_{29}$).}
For $P_{12}$ within the range, $0 \le \lambda < \sqrt{2}$, all eight eigenvalues are strictly negative, making $P_{12}$ the **unique late-time accelerating sink** of the expanding phase space. For $\sqrt{2} < \lambda < \sqrt{6}$, the curvature eigenvalues become positive, turning $P_{12}$ into a saddle. At $\lambda = \sqrt{2}$, four eigenvalues vanish as $P_{12}$ bifurcates with $P_{17}$ and $P_{30}$. Points $P_{13}$ and $P_{29}$ exhibit $\tilde{q} > 0$; while stable within the flat Bianchi~I subspace, positive spatial-curvature eigenvalues render both points saddles in the full Bianchi~IX phase space.

\begin{table}[H]
\tiny
\addtolength{\tabcolsep}{-2.5pt}
\begin{adjustwidth}{-2cm}{-2cm}
\centering
\begin{tabular}{|Sc|Sc|Sc|}
\hline
Point & Eigenvalues & Stability \\

\hline
$P_{12}$ & $\frac{\lambda ^2}{2}-3,\frac{\lambda ^2}{2}-3,\frac{\lambda ^2}{2}-3,\frac{\lambda ^2}{2}-1,\frac{\lambda ^2}{2}-1,\frac{\lambda ^2}{2}-1,\lambda ^2-2,\beta  \lambda +\lambda ^2-3$ & \makecell{Stable for $0\leq\lambda<\sqrt{2}$ \\ Saddle for $\sqrt{2}<\lambda<\sqrt{6}$ \\ Non-hyperbolic at $\lambda=\sqrt2$ (bifurcation with $P_{17}$, $P_{30}$)} \\

\hline
$P_{13}$ & \makecell{$1-\frac{3 \beta }{\beta +\lambda },\frac{\lambda -2 \beta }{2 (\beta +\lambda )},\frac{\lambda -2 \beta }{2 (\beta +\lambda )},\frac{\lambda -2 \beta }{2 (\beta +\lambda )},-\frac{3 (2 \beta +\lambda )}{2 (\beta +\lambda )},-\frac{3 (2 \beta +\lambda )}{2 (\beta +\lambda )},$ \\ $\frac{-3(2\beta+\lambda)\pm i \sqrt{3} \sqrt{16 \beta ^3 \lambda +4 \beta ^2 \left(8 \lambda ^2-15\right)+4 \beta  \lambda  \left(4 \lambda ^2-9\right)+21 \lambda ^2-72}}{4 (\beta +\lambda )}$} & Saddle \\

\hline
\makecell{$P_{14}$\\$P_{15}$\\$P_{16}$} & \makecell{$1-\frac{3\beta}{\beta+\lambda},\frac{3}{4}-\frac{9\beta}{4(\beta+\lambda)},\frac{3}{4}-\frac{9\beta}{4(\beta+\lambda)},-\frac{3(2\beta+\lambda)}{2(\beta+\lambda)},$\\$-\frac{3(2\beta+\lambda)}{4(\beta+\lambda)}\pm\frac{3}{8(\beta+\lambda)^2}\sqrt{-(\beta+\lambda)^2(\Gamma-\sqrt{\Delta})},-\frac{3(2\beta+\lambda)}{4(\beta+\lambda)}\pm\frac{3}{8(\beta+\lambda)^2}\sqrt{-(\beta+\lambda)^2(\Gamma+\sqrt{\Delta})}$} & Saddle \\

\hline
\makecell{$P_{17}$\\$P_{18}$\\$P_{19}$} & \makecell{$12-\frac{216}{\lambda ^2+16},12-\frac{216}{\lambda ^2+16},16-\frac{288}{\lambda ^2+16},\frac{18 (\beta  \lambda -16)}{\lambda ^2+16}+15,6-\frac{144}{\lambda ^2+16},6-\frac{144}{\lambda ^2+16},$ \\ $-\frac{3 \left(-\lambda ^2+\sqrt{\left(\lambda ^2-8\right) \left(13 \lambda ^2-32\right)}+8\right)}{\lambda ^2+16},\frac{3 \left(\lambda ^2+\sqrt{\left(\lambda ^2-8\right) \left(13 \lambda ^2-32\right)}-8\right)}{\lambda ^2+16}$} & \makecell{Saddle for $2<\lambda^2<8$ \\ Non-hyperbolic at $\lambda=\sqrt2$ (coincides with $P_{12}$) \\ and at $\lambda=2\sqrt2$ (coincides with a point of $P_7$)} \\

\hline
\makecell{$P_{20}$\\$P_{21}$\\$P_{22}$} & \makecell{$12-\frac{90}{\beta ^2+8},12-\frac{90}{\beta ^2+8},16-\frac{120}{\beta ^2+8},6-\frac{60}{\beta ^2+8},6-\frac{60}{\beta ^2+8},$ \\ $-\frac{3 \left(-\beta ^2+\sqrt{\left(\beta ^2-2\right) \left(11 \beta ^2+3\right)}+2\right)}{\beta ^2+8},\frac{3 \left(\beta ^2+\sqrt{\left(\beta ^2-2\right) \left(11 \beta ^2+3\right)}-2\right)}{\beta ^2+8},\frac{3 \left(6 \beta ^2+5 \beta  \lambda +8\right)}{2 \left(\beta ^2+8\right)}$} & \makecell{Saddle for $\beta^2<2$ \\ Non-hyperbolic at $\beta=-\sqrt2$ (coincides with a point of $P_7$)} \\

\hline
\makecell{$P_{23}$\\$P_{24}$\\$P_{25}$} & $6,4,3,3,0,0,\pm2i\nu$ & \makecell{Non-hyperbolic, \\ Unstable for all $\nu\ge0$} \\

\hline
\makecell{$P_{26}$\\$P_{27}$\\$P_{28}$} & $6,4,3,3,0,0,\pm\frac{3 i \sqrt{2 \beta ^2+1}}{\beta ^2+2}$ & \makecell{Non-hyperbolic, Unstable \\ (Same as Points $P_{23}$--$P_{25}$ with $\nu=\frac{2\sqrt{2\beta^2+1}}{3(\beta^2+2)}$)} \\

\hline
$P_{29}$ & $\beta ^2-\frac{3}{2},\beta ^2-\frac{3}{2},\beta ^2-\frac{3}{2},\beta ^2+\frac{1}{2},\beta ^2+\frac{1}{2},\beta ^2+\frac{1}{2},2 \beta ^2+1,\beta ^2+\beta  \lambda +\frac{3}{2}$ & \makecell{Saddle for $\beta^2<\frac32$ \\ Non-hyperbolic at $\beta=-\sqrt{\frac32}$} \\

\hline
$P_{30}$ & \makecell{$0,\frac{2 \beta -\lambda }{\sqrt{2}},-\frac{\sqrt{8-3 \lambda ^2}+\lambda }{\sqrt{2}},\frac{\sqrt{8-3 \lambda ^2}-\lambda }{\sqrt{2}},$\\$-\frac{\sqrt{9 \lambda ^2-16}+\lambda }{\sqrt{2}},-\frac{\sqrt{9 \lambda ^2-16}+\lambda }{\sqrt{2}},\frac{\sqrt{9 \lambda ^2-16}-\lambda }{\sqrt{2}},\frac{\sqrt{9 \lambda ^2-16}-\lambda }{\sqrt{2}}$} & \makecell{Saddle for $0<\lambda^2<2$ \\ Merges with $P_{12}$ at $\lambda=\sqrt{2}$} \\

\hline
\end{tabular}
\caption{Eigenvalues and stability of the fixed points $P_{12}$--$P_{30}$ of Table~\ref{table:anisotropic}, for $\lambda\ge0$ and $\beta\le0$, within the domain of existence of each point. Stability is stated only where the point exists ($\tilde Y\ge0$, $\tilde\Omega_{DM}\ge0$ and real $\tilde N_i$). The lines $P_{23}$--$P_{25}$ are parametrized by the common value $\nu$ of their two non-vanishing curvature variables; $P_{26}$--$P_{28}$ are particular points on these lines. The eigenvalues of $P_{30}$ are computed with respect to the $D$-normalized time, which coincides with the $H$-normalized time on the shell $\tilde H=1$. For $P_{14}$--$P_{16}$,
$\Gamma=-48+12\beta^3\lambda+17\lambda^2+10\beta\lambda(\lambda^2-3)+\beta^2(22\lambda^2-72)$ and
$\Delta=64(\beta^2+2)(\beta^2+18)-32\beta(6\beta^4+19\beta^2-42)\lambda+4(36\beta^6-220\beta^4-555\beta^2-248)\lambda^2+4\beta(132\beta^4-180\beta^2-323)\lambda^3+(724\beta^4+164\beta^2+121)\lambda^4+220\beta(2\beta^2+1)\lambda^5+100\beta^2\lambda^6$.}
\label{eigenvalue_anisotropic2}
\end{adjustwidth}
\end{table}

\paragraph{Single-curvature Bianchi II points ($P_{14}$--$P_{22}$).}
All nine Bianchi II states decelerate ($\tilde{q} > 0$). Their two transverse curvature eigenvalues, given by $\frac{3}{2}\tilde{q}$, are strictly positive, identifying $P_{14}$--$P_{22}$ as transient saddles. These triplets mark bifurcation boundaries: $P_{14}$--$P_{16}$ link $P_{13}$ to the curvature sector, $P_{17}$--$P_{19}$ branch from $P_{12}$ at $\lambda^2=2$ and join the Jacobs sphere at $\lambda^2=8$, while $P_{20}$--$P_{22}$ intersect the Jacobs sphere at $\beta^2=2$.

\paragraph{Double-curvature Bianchi VII$_0$ lines ($P_{23}$--$P_{28}$).}
The eigenvalues are $\{6, 4, 3, 3, 0, 0, \pm 2i\nu\}$ are independent of $\lambda$. Four strictly positive eigenvalues render $P_{23}$--$P_{28}$ highly unstable. The imaginary pair $\pm 2i\nu$ governs shear-curvature oscillations ($\nu = 1$ for $P_{23}$--$P_{25}$, and $\nu = \frac{3\sqrt{2\beta^2+1}}{2(\beta^2+2)}$ for the Taub-like set $P_{26}$--$P_{28}$).

\paragraph{Closed isotropic point ($P_{30}$).}
For $0<\lambda^2<2$, $P_{30}$ has exactly one positive eigenvalue, $(\sqrt{8-3\lambda^2}-\lambda)/\sqrt2$, one zero eigenvalue, and six eigenvalues with negative real parts. It is a saddle point. This separates expanding Bianchi~IX trajectories that isotropize towards $P_{12}$ from those that recollapse \cite{vandenHoogen1997}. At $\lambda^2=2$ the positive eigenvalue vanishes and $P_{30}$ merges with $P_{12}$.

\bibliographystyle{ieeetr}
\bibliography{ref}

\end{document}